\documentclass[aps,10pt,prd,superscriptaddress,preprintnumbers,%
showkeys,nofootinbib,showpacs, fleqn]{revtex4-2}
\usepackage{amsmath,amsfonts,amssymb,amscd,amsxtra,amsthm}
\usepackage{graphicx}  
\usepackage{epstopdf}
\usepackage{dcolumn}  
\usepackage{bm}          
\usepackage{slashed}
\usepackage{cancel}
\usepackage{float}
\usepackage{mathtools}
\usepackage{amsbsy}
\usepackage{amstext}
\usepackage{hyperref}

\usepackage{wasysym}
\usepackage[utf8]{inputenc} 
\usepackage{booktabs} 
\usepackage[normalem]{ulem} 
\usepackage[dvipsnames]{xcolor} 
\usepackage{tabularx}
\usepackage{enumitem}  
\usepackage{array} 
\usepackage{slashed}
\usepackage{tikz}
\usepackage{float}
\usepackage{multirow}
\usepackage{cancel}
\usepackage{physics}
\renewcommand\sout{\bgroup \color{red} \ULdepth=-.5ex \ULset}

\makeatletter

\begin{document}
\preprint{INHA-NTG-01/2025}
\title{Nucleon and singly heavy baryons from the QCD instanton vacuum}    

\author{Yongwoo Choi}
\email{sunctchoi@gmail.com}
\affiliation{Center for Extreme Nuclear Matters (CENuM), Korea
  University, Republic of Korea} 
\affiliation{Department of Physics, Inha University, Incheon 22212,
  Republic of Korea} 

\author{Hyun-Chul Kim}
\email{hchkim@inha.ac.kr}
\affiliation{Department of Physics, Inha University, Incheon 22212,
  Republic of Korea} 
\affiliation{School of Physics, Korea Institute for Advanced Study
  (KIAS), Seoul 02455, Republic of Korea} 

\date{\today}

\begin{abstract}
We construct an effective chiral theory for the nucleon, based on the 
low-energy effective QCD partition function from the QCD instanton
vacuum. We fully consider the momentum-dependent dynamical quark mass
whose value at the zero virtuality of the quark is determined by the
gap equation from the instanton vacuum, $M_0=359$ MeV. The nucleon
emerges as a state of $N_c$ valence quarks bound by the pion mean
field, which was created self-consistently by the $N_c$ valence
quarks. In the large Euclidean time, the classical nucleon mass is
evaluated by minimizing the sum of the $N_c$ discrete-level energies
and the Dirac-continuum energy: $M_{\text{cl}}=1.2680$ GeV. The pion
mean-field solution turns out broader than the local chiral
quark-soliton model. The zero-mode quantization furnishes the nucleon
with proper quantum numbers such as the spin and isospin. We compute
the moment of inertia $I=1.3853$ fm by using the self-consistent
mean-field solution, which yields the $\Delta -N$ mass splitting 
$M_{\Delta-N} =213.67$ MeV. In the same manner, singly heavy baryons
can be described as a bound state of the $N_c-1$ valence quarks with
the corresponding pion mean field, with the heavy quark regarded as a
static color source. The mass splitting of the singly heavy 
baryons is obtained to be $M_{\Sigma_Q-\Lambda_Q}=206.20$ MeV, which
are in good agreement with the experimental data. The effective chiral
theory developed in the present work will provide a solid theoretical
framework to investigate gluonic observables of both the light and
singly heavy baryons. 
\end{abstract}

\maketitle


\section{Introduction}\label{sec:1}
Since Rutherford christened the simplest nucleus of the hydrogen atom
as the proton~\cite{Masson:1920}, the structure of the nucleon has
been one of the central issues in physics for over a century.
The possibility of an internal structure was first suspected when the
magnetic dipole moments of the proton and neutron were
measured~\cite{Frisch:1933, Estermann:1933, Estermann:1937,
  Alvarez:1940zz}. As electrons probed the
electromagnetic structure of the proton~\cite{Hofstadter:1956qs},
including its charge radius, the first internal picture of the proton
emerged: a bare nucleon encompassed by a pion
cloud~\cite{Yennie:1957skg}. However, electron-proton deep inelastic
scattering revealed that partons constitute the proton, carrying
fractions of its momentum~\cite{Breidenbach:1969kd,
  Bjorken:1969ja}. With the advent of quantum chromodynamics (QCD),
these partons were identified as quarks and
gluons~\cite{Fritzsch:1973pi, Gross:2022hyw}. Since then, numerous
models and theories for the nucleon have been proposed, including  
the quark potential models~\cite{DeRujula:1975qlm}, the MIT bag
model~\cite{Chodos:1974pn}, the Skyrme models~\cite{Skyrme:1961vq,
  Adkins:1983ya}, the chiral bag model~\cite{Thomas:1982kv}, the Goldstone
boson-exchange model~\cite{Glozman:1997ag}, the Nambu--Jona-Lasinio
(NJL) model~\cite{Weigel:1992sf, Alkofer:1994ph}, nonlocal NJL
model~\cite{Golli:1998rf, Broniowski:2001cx}, Dyson-Schwinger
approaches~\cite{Eichmann:2009qa}, QCD sum
rules~\cite{Shifman:1978bx,Shifman:1978by,  Ioffe:1983ju,
  King:1986wi}, lattice
QCD~\cite{Leinweber:2003dg,QCDSF-UKQCD:2003hmh,Alexandrou:2017oeh}
and so on. 

While these approaches have their strengths and weaknesses in
describing the nucleon, a modern view of nucleon structure,
represented by generalized parton distributions
(GPDs)~\cite{Muller:1994ses, Ji:1996nm, Radyushkin:1996nd} and 
gravitational form factors~\cite{Polyakov:2018rew}, unveils a
significant role of gluons in understanding the fundamental properties
of the nucleon such as its mass, spin, and
$D$-term~\cite{Polyakov:1999gs}. The upcoming electron-ion   
collider (EIC) at the Brookhaven National Laboratory (BNL) is expected
to disclose the physics of gluons inside a
nucleon~\cite{AbdulKhalek:2021gbh}.  However, there are few
theoretical frameworks to deal with gluons, particularly considering
their nonperturbative aspect inside a nucleon. While lattice QCD
simulates various gluonic observables often represented by
higher-twist operators, one still requires firm theoretical grounds to
interpret and analyze the physics of nonperturbative gluons. 

The QCD instanton vacuum provides a theoretical framework for the 
realization of the spontaneous breakdown of chiral symmetry
(SB$\chi$S)~\cite{Shuryak:1981ff, Diakonov:1983hh,
  Diakonov:1985eg} (see also the reviews~\cite{Schafer:1996wv,
  Diakonov:2002fq}). The instanton ensemble in the dilute liquid
approximation is characterized by the average size of the instanton
$\bar{\rho}$ and the interdistance between instantons $\bar{R}$, which
are approximately given by $\bar{\rho}\approx 1/3$ fm and
$\bar{R}\approx 1$ fm if one uses
$\Lambda_{\mathrm{QCD}}^{\overline{MS}} = 280$
MeV~\cite{Shuryak:1981ff, Diakonov:1983hh, Diakonov:2002fq}.
These values of $\bar{\rho}$ and $\bar{R}$ yield the proper values of  
the gluon condensate~\cite{Shifman:1978bx} and of the
topological susceptibility of the vacuum~\cite{Diakonov:1995qy}.
The instanton vacuum induces an effective nonlocal $2N_f$ quark-quark
interaction, where $N_f$ is the number of flavors. After bosonization,
one can derive the effective chiral action with a momentum-dependent
dynamical quark mass. Based on this action, an effective chiral theory
of the nucleon was developed in Ref.~\cite{Diakonov:1987ty}. It was
further elaborated and extended to the SU(3) case~\cite{Blotz:1992pw},
which was called the chiral quark-soliton model ($\chi$QSM) (see also 
following reviews~\cite{Christov:1995vm, Diakonov:1997sj}). An
effective chiral theory of the nucleon was motivated by Witten's
seminal papers~\cite{Witten:1979kh, Witten:1983tw, Witten:1983tx}. In
the large $N_c$ (the number of colors) limit, the mass of the
nucleon is proportional to $N_c$, while meson fluctuations are
suppressed by $1/N_c$. Thus, to the leading order in the $1/N_c$
expansion, the presence of $N_c$ valence quarks creates an effective 
pion mean field emerging from vacuum polarization. This
mean field influences the $N_c$ valence quarks in a self-consistent
manner, allowing the nucleon to be viewed as a state of 
$N_c$ valence quarks bound by the pion mean field. 

In the $\chi$QSM, the momentum dependence of the dynamical quark mass 
was typically turned off for simplicity, with a specific
regularization scheme introduced. The regularization scheme must 
only apply to the real part of the effective chiral action (E$\chi$A),
while the imaginary part remains intact. Otherwise, usual
regularization schemes such as the proper-time and Pauli-Villars
regularizations would spoil the chiral anomaly related to the
Wess-Zumino-Witten (WZW) term arising from the derivative expansion of
the imaginary part of the E$\chi$A. When keeping the momentum
dependence of the dynamical quark mass, the chiral anomaly is safely
kept unchanged~\cite{Ball:1993ak}. The momentum-dependent dynamical
quark mass from the instanton vacuum plays the role of a natural regulator.   
Moreover, it arises from the fermionic zero-mode solutions, and
encodes the nonperturbative quark-gluon dynamics via the instanton
medium~\cite{Diakonov:1985eg, Diakonov:2002fq}. Thus, it is crucial to
keep the momentum dependence of the dynamical quark mass in describing
the structure of hadrons. This is especially important when
computing the baryonic matrix elements of effective QCD gluon
operators, for which the momentum-dependent quark 
mass comes into an essential role. In Ref.~\cite{Diakonov:1995qy}, it
was shown that gluon operators can be expressed in terms of effective
quark operators. They can further be written in terms of the quark and 
chiral fields~\cite{Balla:1997hf}. Consequently, to compute the
baryonic matrix elements of the gluon operators, it is necessary to
develop an effective chiral theory of the nucleon with the
momentum-dependent quark mass from the instanton vacuum.   

This work aims to construct an effective chiral theory for the nucleon
from the instanton vacuum.  In Section~\ref{sec:2}, we begin by
recapitulating how the low-energy effective QCD partition function is
derived from the instanton vacuum. Section~\ref{sec:3} demonstrates
how the valence quarks are self-consistently bound by the pion mean
fields, which emerge from the vacuum polarization. We compute the
nucleon correlation function, consisting of $N_c$ valence quarks, 
so that we evaluate the valence-quark (discrete-level)
and sea-quark (Dirac continuum) energies. By solving the
equations of motion self-consistently, we derive the pion mean-field
solution and the mass for the classical nucleon. In
Section~\ref{sec:4}, we show that the baryon number is solely
determined by the $N_c$ valence quarks by computing the three-point
correlation function with the baryon current, with the gauge
invariance kept intact. Section~\ref{sec:5} is dedicated to the
zero-mode quantization, which furnishes the nucleon with the proper
quantum numbers such as spin and isospin. We discuss the results for
the $\Delta$ isobar and nucleon mass splitting. In
Section~\ref{sec:6}, we extend the current effective theory to the
description of singly heavy baryons as a bound state of the $N_c-1$
valence quarks. The final section summarizes the main points of the
current work, and gives an outlook in the future. 

\section{effective chiral action from the instanton vacuum
  \label{sec:2}} 
In Euclidean space, we start from the effective low-energy QCD partition
function~\cite{Diakonov:1985eg, Diakonov:2002fq} with the number of
flavors $N_f=2$, expressed as  
\begin{align}
    \mathcal{Z}_{\mathrm{eff}}
  &=\int\mathcal{D}\psi^{\dagger}\mathcal{D} \psi~{\rm 
         exp}\Big[-\sum_{f=1}^{2}\int d^{4}x~
\psi^{\dagger}_{f}(x)(-i\slashed{\partial})\psi_{f}(x)\Big]~
\bigg(\frac{Y_{2}^{+}}{VM_{1}^{N_{f}}}\bigg)^{N_{+}}
\bigg(\frac{Y_{2}^{-}}{VM_{1}^{N_{f}}}\bigg)^{N_{-}}, 
\label{eq:1}  
\end{align}
where the mass factor $M_{1}$ separates the 
high-frequency part of the fermionic determinant from 
the low-frequency part~\cite{Diakonov:1984vw}.
The high-frequency part are integrated out and its effects are
absorbed into the strong coupling constant in a renormalization group 
sense~\cite{Diakonov:1985eg}. The matching was carried out
between the high- and low-frequency constributions to the full
fermionic determinant in the instanton vacuum~\cite{Diakonov:1984vw}. 
$N^\pm$ denote the number of the instantons ($I$'s) and anti-instantons
($\bar{I}$'s) in the four-dimensional volume $V$. Averaging over the
positions and color orientations of the instantons and anti-instantons,
we can show that the instanton vacuum induces the $2N_f$ 
vertex interactions. Thus, $Y_{2}^{+}$ ($Y_2^{-}$) in
Eq.~\eqref{eq:1} represent the instanton-induced
(antiinstanton-induced) four-quark interactions for  
$N_{f}=2$~\cite{Diakonov:1997sj, Diakonov:2002fq}  
\begin{align}
     Y_{2}^{+} =&~ \frac{i^2}{N_c^2-1} \int \frac{d^4k_1}{(2\pi)^{4}}
     \frac{d^4k_2}{(2\pi)^{4}}\frac{d^4l_1}{(2\pi)^{4}}
     \frac{d^4l_2}{(2\pi)^{4}}~\delta (k_1 + k_2 - l_1
                   -l_2)  
 \int d\rho~ \nu(\rho) (2\pi\rho)^4 F(k_1\rho)F(k_2\rho)
                   F(l_1\rho) F(l_2\rho)  \cr
&\times \frac1{2!} \epsilon^{f_1 f_2} \epsilon_{g_1g_2} \left[
\left(1-\frac1{2N_c}\right) \left \{\psi_{f_1}^\dagger (k_1) P_L \psi^{g_1}
  (l_1)\right\} \left\{\psi_{f_2}^\dagger (k_2) P_L \psi^{g_2}
  (l_2)\right\}  \right. \cr
&+ \left. \frac1{8N_c} \left \{\psi_{f_1}^\dagger (k_1) P_L
  \sigma_{\mu\nu} \psi^{g_1} (l_1)\right\} \left \{\psi_{f_2}^\dagger
  (k_2) P_L \sigma_{\mu\nu} \psi^{g_2} (l_2)\right\} \right],\nonumber 
\end{align}
\begin{align}
  Y_{2}^{-} =&~\frac{i^2}{N_c^2-1} \int \frac{d^4k_1}{(2\pi)^{4}}
  \frac{d^4k_2}{(2\pi)^{4}}\frac{d^4l_1}{(2\pi)^{4}}
  \frac{d^4l_2}{(2\pi)^{4}}~\delta (k_1 + k_2 - l_1
                   -l_2)  
 \int d\rho~ \nu(\rho) (2\pi\rho)^4 F(k_1\rho)F(k_2\rho)
                   F(l_1\rho) F(l_2\rho)  \cr
&\times \frac1{2!} \epsilon^{f_1 f_2} \epsilon_{g_1g_2} \left[
\left(1-\frac1{2N_c}\right) \left \{\psi_{f_1}^\dagger (k_1) P_R \psi^{g_1}
  (l_1)\right\} \left\{\psi_{f_2}^\dagger (k_2) P_R \psi^{g_2}
  (l_2)\right\}  \right. \cr
&+ \left. \frac1{8N_c} \left \{\psi_{f_1}^\dagger (k_1) P_R
  \sigma_{\mu\nu} \psi^{g_1} (l_1)\right\} \left \{\psi_{f_2}^\dagger
  (k_2) P_R \sigma_{\mu\nu} \psi^{g_2} (l_2)\right\} \right],  
\label{eq:2}
\end{align}
where $\nu(\rho)$ denotes the size distribution of the
instanton. We consider the average of instanton size by replacing
$\rho$ by its average value $\bar{\rho}$, i.e., $\nu(\rho)
=\delta(\rho-\bar{\rho})$. This is plausible, since the width of the
instanton size distribution is of the order $\mathcal{O}(1/N_c)$ in
the dilute liquid approximation of the instanton
medium~\cite{Kim:2005jc}. $P_L$ and $P_R$ represent the projection
operators defined by $P_L=(1-\gamma_5)/2$ and $P_R=(1+\gamma_5)/2$.  
The summation over $N_{\pm}$ in Eq.~(\ref{eq:1})
are performed by the saddle-point approximation to average the
ensemble of $N_{+}$ $I$'s and $N_{-}$ $\bar{I}$'s in the
thermodynamic limit, i.e., $N_{\pm}\rightarrow\infty$,
$V\rightarrow\infty$, with $N_{\pm}/V$ fixed.  
Since the strong interaction is $CP$-invariant, we set $N_{\pm}=N/2$.    

\begin{figure}[h!]\centering
\includegraphics[width=0.6\columnwidth]{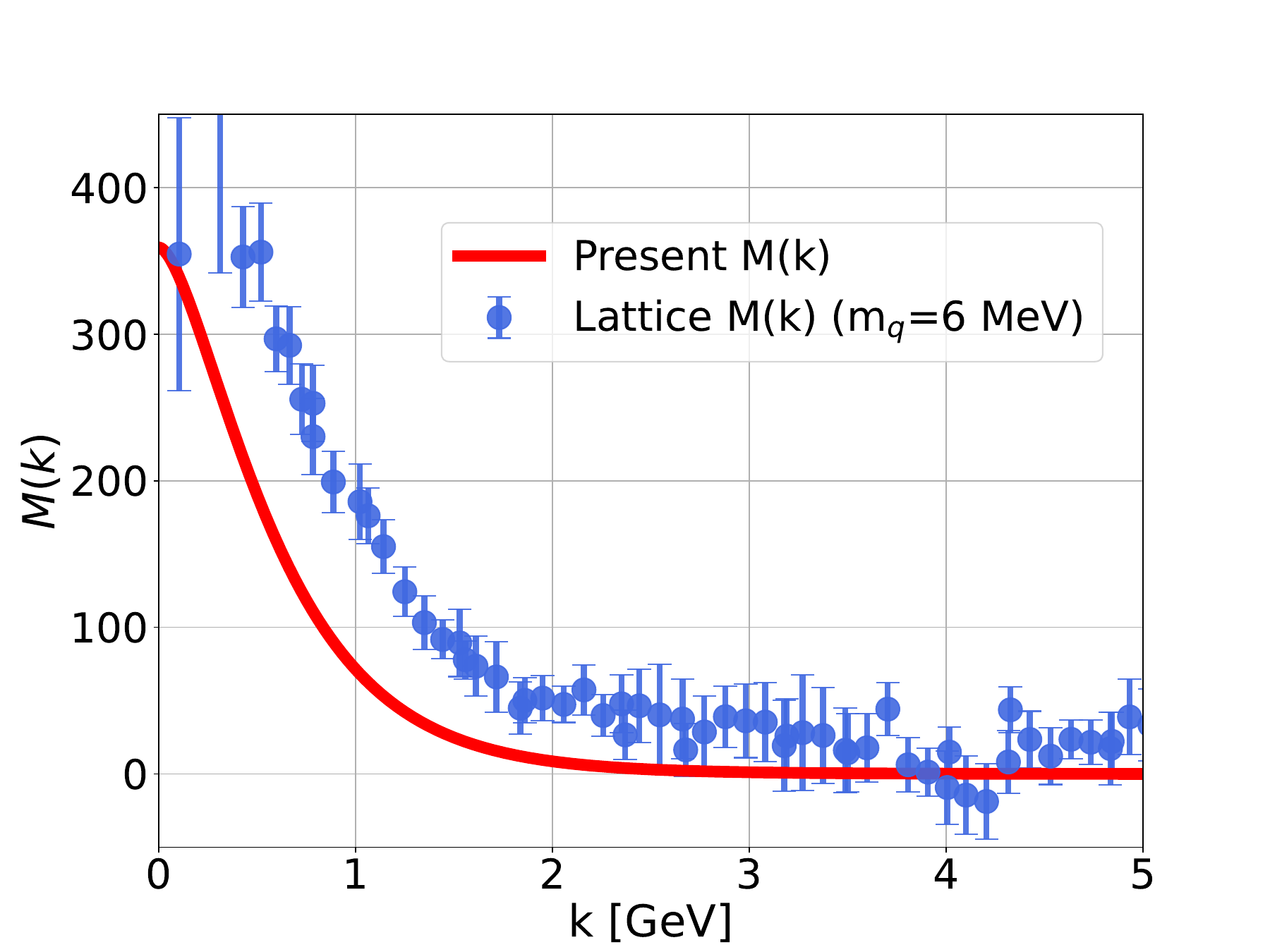}
\caption{\label{fig:1}%
Momentum-dependent dynamical quark mass $M(k)$ as a function of $k$. 
The points are lattice calculation with the current quark mass $m_{q}=6~\mathrm{MeV}$.}
\end{figure}

Taking the chiral limit ($m_{f}=0$) and using the saddle-point
approximation, we obtain the gap equation as 
\begin{align}
    4 N_{c}\int\frac{d^{4}k}{(2\pi)^{4}}\frac{[M(k)]^{2}}{k^{2}+[M(k)]^{2}}
  =\frac{N}{V}={\bar{R}}^{-4},
\label{eq:3}  
\end{align}
where $\bar{R}$ stands for the average distance between instantons and 
$M(k)$ represents the momentum-dependent quark mass defined as 
\begin{align}
    M(k)=M_{0}F(k\bar{\rho})^{2},
\label{eq:4}
\end{align}
where the quark form factor $F(k)$ is the Fourier transform of
the fermionic zero-mode solution in the presence of the instanton in
the singular gauge~\cite{tHooft:1976snw}, expressed as 
\begin{align}
F(k\bar{\rho})=-z~\frac{d}{dz}\Big[I_{0}(z)K_{0}(z)-I_{1}(z)K_{1}(z)\Big] 
\Big{\vert}_{z=k\bar{\rho}/2}.
\label{eq:5}
\end{align}
$M_0$ is the dynamical quark mass at the zero virtuality of the quark,
which is determined by the gap equation: $M_0=359~\mathrm{MeV}$.  
In Fig.~\ref{fig:1} we compare the numerical result for
Eq.~\eqref{eq:5} with those from the lattice
data~\cite{Leinweber:2003dg}. Note that the momentum-dependent quark 
mass is a gauge-dependent quantity, and the lattice data employs a
different gauge. Thus, the present form of $M(k)$ does not necessarily
reproduce the absolute values of the lattice data. A general behavior
of $M(k)$ is in line with the lattice data.

Taking the large $N_c$ limit, we can ignore the tensor interactions in
Eq.~\eqref{eq:2}. Thus, the quark-quark interactions $Y_2^{\pm}$ can
be written as 
\begin{align}
Y_2^{\pm} & = \left(\frac1{N_c}\right)^2 \int d^4 x~
            \mathrm{det}[iJ^\pm(x)]  
\label{eq:6}
\end{align}
with 
\begin{align}
J_{fg} (x) = \int \frac{d^4 k}{(2\pi)^4}\frac{d^4 l}{(2\pi)^4}~ e^{i(k-l)\cdot x}
  (2\pi\bar{\rho} F(k\bar{\rho})) (2\pi\bar{\rho} F(l\bar{\rho}))    
\left[ \psi_{f}(k) \frac{1\mp \gamma_5}{2}
  \psi_{g}(l)\right].
\label{eq:7}  
\end{align}
After bosonization, the effective low-energy QCD partition
function is expressed as 
\begin{align}
\mathcal{Z}_{\mathrm{eff}}[\psi^\dagger,\psi, \pi^a] =& \int
  \mathcal{D}\pi^a \int  \mathcal{D}\psi^\dagger  \mathcal{D}\psi ~\exp\left[ 
-\int d^4 x \left\{ \psi_f^\dagger (x) (-i\rlap{/}{\partial}) \psi^f (x)
   - i M_0 \int \frac{d^4k}{(2\pi)^4}\frac{d^4l}{(2\pi)^4}~ F(k)F(l)
\right. \right. \cr 
& \times \left. \left. \psi_f^{\dagger}(x)  
  \left(U_g^f(x) P_L + U_g^{\dagger f}(x) P_R\right)
    \psi^g(x)    \right\}\right],
\label{eq:8}  
\end{align}
where $\pi^a$ represent the Nambu-Goldstone (NG) boson
fields. The chiral field $U(x)$ is defined as $U (x) := \exp[i\pi^a(x)
\tau^a]$ with $a=1,~2,~3$. In coordinate space, we rewrite it as 
\begin{align}
\mathcal{Z}_{\mathrm{eff}}= \int
  \mathcal{D}\pi^a \int  \mathcal{D}\psi^\dagger  \mathcal{D}\psi ~\exp\left[ 
-\int d^4 x \left\{ \psi_f^\dagger (x) (-i\rlap{/}{\partial}) \psi^f (x)
   - i M_0  \psi_f^{\dagger} (x) \overleftarrow{F}(i\partial)
  (U^{\gamma_5})_g^f (x) 
    \overrightarrow{F}(i\partial)\psi^g(x)  \right\}\right],   
\label{eq:9}   
\end{align}
where $U^{\gamma_5}$ is written as 
\begin{align}
(U^{\gamma_5})_g^f (x) = (U)_g^f(x) P_L + (U^{\dagger})_g^{f}(x) P_R.  
\label{eq:10}
\end{align}
After integrating over the
quark fields, we obtain the following partition function
\begin{align}
Z_{\mathrm{eff}}[\pi^a] =  \int \mathcal{D}\pi^a ~\mathrm{Det}\left[
  i\rlap{/}{\partial} +   i M_0 
  \overleftarrow{F}(i\partial) U^{\gamma_5}[\pi^a(x)]  
    \overrightarrow{F}(i\partial)   \right] 
\label{eq:11}
\end{align}
Since pionic quantum fluctuations are suppressed by $1/N_c$, we can
ignore it. So, we can perform the integration by the saddle-point
approximation~\cite{Witten:1979kh, Witten:1983tw,
  Witten:1983tx}. Note that the zero-modes around the classical
solution must be considered completely, which we will discuss in 
Section~\ref{sec:5}. Thus, we obtain the nonlocal effective
chiral action (NE$\chi$A) as follows: 
\begin{align}
S_{\mathrm{eff}}[U]= -N_c \mathrm{Tr}\log (D[U]),
\label{eq:12}
\end{align}
where the Dirac differential operator $D[U]$ is given by
\begin{align}
  D [U] =  i\rlap{/}{\partial} + i M_0  \overleftarrow{F}(i\partial)
  U^{\gamma_5} (x) \overrightarrow{F}(i\partial).   
\label{eq:13}
\end{align}
We take this NE$\chi$A as the starting point for the current effective
chiral theory for the nucleon. Note that Eq.~\eqref{eq:12} contains
all orders of effective chiral Lagrangians with the renormalized
effective low-energy constants (LECs), which can be shown by the
derivative expansion~\cite{Choi:2003cz}. 

The pion decay constant $f_\pi$ is defined in terms of the transition
matrix element 
\begin{align}
\langle 0 | A_\mu^a(x) |\pi^b(p)\rangle = i f_\pi(p^2) p_\mu
\delta^{ab} e^{ip\cdot x},
\label{eq:14}
\end{align}
where $A_\mu^a(x)$ is the axial-vector current defined by
$A_\mu^a(x):=-i\psi^{\dagger}(x) \gamma_\mu \gamma_5 \tau^a \psi(x)$ in usual sense.
Since the vector and axial-vector currents are not conserved in the
presence of the momentum-dependent quark mass because of its nonlocal 
interaction. Without restoring the gauge invariance corresponding to
the axial-vector current, one cannot get the correct expression for
the pion decay constant~\cite{Pagels:1979hd, Holdom:1990iq,
  Bowler:1994ir, Golli:1998rf, Broniowski:2001cx}. We will discuss the
problem of the current conservation in the presence of the nonlocal
interactions, when we compute the baryon number of the classical
nucleon. The pion decay constant is also regarded as one of the LECs,
because the effective chiral Lagrangian to the leading order in the
derivative expansion is just the Weinberg term that is a kinetic
Lagrangian for the pion:
\begin{align}
\mathcal{L}_{\mathrm{eff}}^{\mathcal{O}(p^2)} = \frac{f_\pi^2}{4}
  \mathrm{Tr}(\partial_\mu U \partial_\mu U^\dagger),   
\label{eq:15}
\end{align}
where $f_\pi^2$ is shown to be expressed as~\cite{Choi:2003cz}: 
\begin{align}
f_{\pi}^{2}=
4 N_{c}\int\frac{d^{4}k}{(2\pi)^{4}}\frac{M^{2}-\frac{1}{2}\vert
k\vert M M' + \frac{1}{4} \vert k\vert^{2}{M'}^{2}}{(k^{2}+M^{2})^{2}}
  \mbox{ with } M'=\frac{dM(k)}{dk}.
\label{eq:16}
\end{align}
Equation~\eqref{eq:16} is exactly the same as that obtained in
Refs.~\cite{Bowler:1994ir, Golli:1998rf, Broniowski:2001cx}. 
With the values of $\bar{\rho} = 1/3$ fm and $\bar{R}=1$ fm, 
we obtain $f_\pi= 90.4~ \mathrm{MeV}$. The value of $f_\pi$
in chiral perturbation theory is determined to be $f_\pi=92.2\pm 0.2$
MeV~\cite{Donoghue:2022wrw}. In the chiral limit, it becomes smaller
than that with the finite pion mass~\cite{Gasser:1983yg,
  Gasser:1984gg}.

\section{Classical nucleon from the instanton vacuum \label{sec:3}} 
\begin{figure}[htp]\centering
\includegraphics[width=0.45\columnwidth]{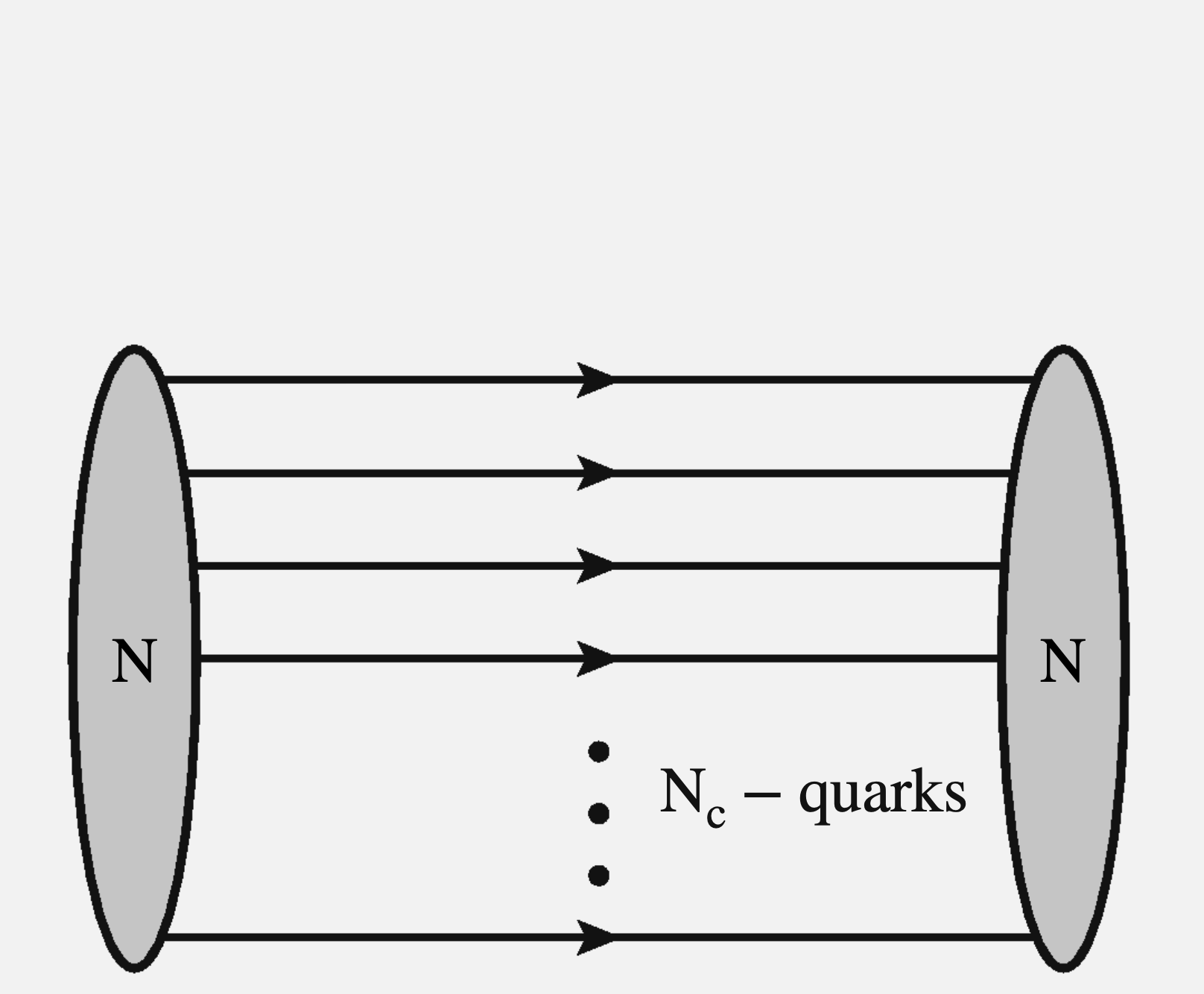}\;\;\;
\includegraphics[width=0.45\columnwidth]{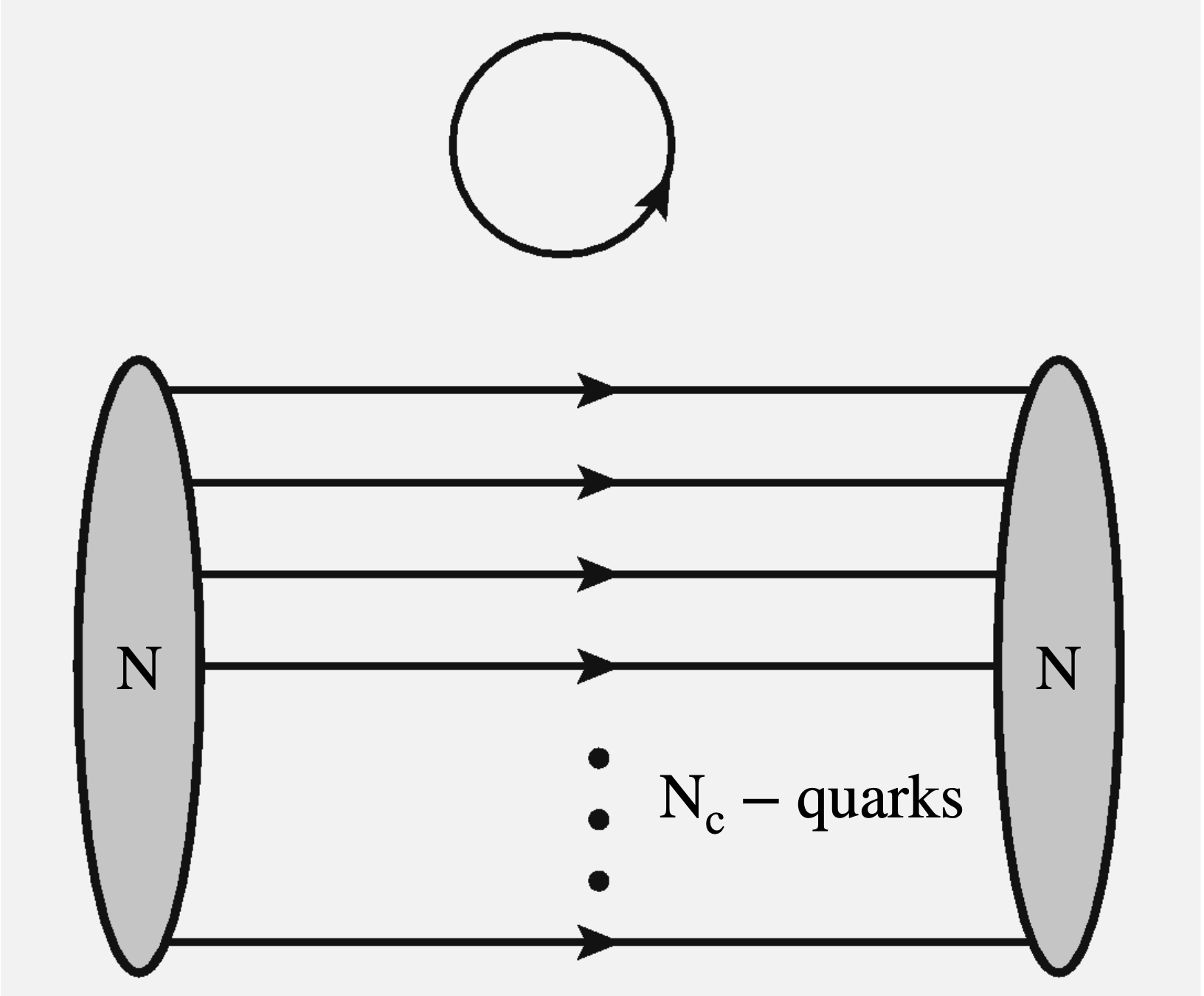}
\caption{\label{fig:2}%
Nucleon correlation function consisting of $N_{c}$ valence quarks in
the background pion field. The left panel depicts the valence-quark
(discrete-level) contribution, whereas the right panel draws the
sea-quark (Dirac-continuum) contribution.}  
\end{figure}

In the large $N_{c}$ limit, a baryon can be viewed as a bound state of
$N_{c}$ valence quarks, which polarize the vacuum and create the pion
mean field~\cite{Witten:1979kh}. The $N_c$ valence quarks are bound by
the pion mean field in a self-consistent manner. Figure~\ref{fig:2}
illustrates the nucleon correlation function. The left panel of
Fig.~\ref{fig:2} depicts the valence-quark contribution, whereas the
right panel draws the sea-quark one.   
The nucleon state is expressed in terms of the Ioffe-type nucleon
current \( J_{N}(x) \) in Euclidean space (\( x_{0}=-ix_{4} \)): 
\begin{align}
| N(p)\rangle & =  \lim _{x_{4}\rightarrow -\infty }
e^{ip_{4}x_{4}}\mathcal{N}(\bm{p})\int d^{3}x~
e^{i\bm{p}\cdot \bm{x}} J^{\dagger }_{N}(\bm{x}, x_4)|0\rangle \cr
\langle N(p')| & =  \lim _{y_{4}\rightarrow +\infty }
e^{-i{p}_{4}' y_{4}}\mathcal{N}^*(\bm{p}')\int d^{3}y~
e^{-i\bm{p'}\cdot \bm{y}}\langle 0|J_{N}(\bm{y}, y_4),
\label{eq:17}
\end{align}
where $\mathcal{N}(\bm{p})$ and $\mathcal{N}^*(\bm{p}')$ denote the
normalization constants. $J_N$ and $J_N^\dagger$ represent the
Ioffe-type current consisting of the $N_c$ valence
quarks~\cite{Diakonov:1987ty}, defined by  
\begin{align}
    J_{N}(x)&=\frac{1}{N_{c}!}\epsilon^{\alpha_{1}\cdots\alpha_{N_{c}}}
              \Gamma_{(TT_3)(SS_3)}^{f_{1},\cdots,f_{N_{c}}}\psi_{\alpha_{1}f_{1}}(x)~
              \cdots~\psi_{\alpha_{N_{c}}f_{N_{c}}}(x), \cr
    J_{N}^{\dagger}(y)&=\frac{1}{N_{c}!}
  \epsilon^{\beta_{1}\cdots\beta_{N_{c}}}
  (\Gamma_{(TT_3)(SS_3)}^{g_{1},\cdots,g_{N_{c}}})^{*}
  (-i\psi^{\dagger}_{\beta_1 g_1}\gamma_{4})
  (y)~\cdots ~
  (-i\psi^{\dagger}_{\beta_{N_{c}}g_{N_{c}}}\gamma_{4})(y),
\label{eq:18}
\end{align} 
where $\alpha_i$ and $\beta_i$ denote color indices configuring the
color-singlet baryon, $f_i$ and $g_i$ are flavor indices, and
$\Gamma^{f}$ is a symmetric matrix that carries the spin and isospin
of nucleon. The nucleon state is normalized as $\langle N (p',S_3')|
N(p,S_3)\rangle = 2p_0 \delta_{S_3'S_3} (2\pi)^3
\delta^{(3)}(\bm{p}'-\bm{p})$. In the large $N_c$ limit, it is reduced
to be $\langle N (p',S_3')| N(p,S_3)\rangle = 2M_N \delta_{S_3'S_3} (2\pi)^3
\delta^{(3)}(\bm{p}'-\bm{p})$, where $M_N$ is the nucleon mass. 
The normalization of the nucleon state is explicitly checked by
computing $\langle N (p',S_3')| N(p,S_3)\rangle$ as
\begin{align}
\langle N (p',S_3')| N(p,S_3)\rangle =&~
\frac{1}{\mathcal{Z}_{\mathrm{eff}}[0]}
  \mathcal{N}^*(p')\mathcal{N}(p)
\lim_{x_4\to -\infty}  \lim_{y_4\to \infty}   
\exp\left(-ip_4' y_4+ip_4 x_4\right)  \cr
& \times \int  d^3x d^3y~ 
  \exp(-i\bm{p}'\cdot \bm{y}+  i\bm{p}\cdot \bm{x})
 \int \mathcal{D}\psi^\dagger \mathcal{D} \psi \mathcal{D} U 
~J_N (y) J_N^\dagger(x) \cr
&\times \exp\left[-\int d^4z\left\{
  (\psi^\dagger(z))_{\alpha}^{f} \left( -i\rlap{/}{\partial}  - i
  M_0 \overleftarrow{F}(i\partial) U^{\gamma_5}
  \overrightarrow{F}(i\partial) \right)_{fg} \psi^{g\alpha}(z)
  \right\} \right].
\label{eq:19}
  \end{align}
Taking the large $N_c$ limit and large Euclidean time
$y_4-x_4=\mathcal{T}\to\infty$, we have $-ip_4'=-ip_4=M_N\sim
\mathcal{O}(N_c)$. Thus, we obtain   
  \begin{align}
\langle N (p',S_3')| N(p,S_3)\rangle =&~
 \frac{1}{\mathcal{Z}_{\mathrm{eff}}[0]}\mathcal{N}^*(p')\mathcal{N}(p) 
 \exp\left(-M_N \mathcal{T}\right) \cr
& \times \int  d^3x d^3y ~
  \exp(-i\bm{p}'\cdot \bm{y}+  i\bm{p}\cdot \bm{x}) \langle 0|
J_N (\bm{y}, +\mathcal{T}/2)
  J_N^\dagger(\bm{x},-\mathcal{T}/2)|0\rangle,  
\label{eq:20}   
\end{align}
where we define the nucleon correlation function as  
\begin{align}
  \Pi_{N}(\mathcal{T})&=\langle 0| J_{N}(\bm{0},+\mathcal{T}/2)
  J_{N}^{\dagger}(\bm{0},-\mathcal{T}/2)|0\rangle\cr
  &=\frac{1}{\mathcal{Z}_{\mathrm{eff}}[0]}
  \int\mathcal{D}\psi^{\dagger} 
  \mathcal{D}\psi\mathcal{D}U~ 
  J_{N}(\bm{0},+\mathcal{T}/2) J_{N}^{\dagger}(\bm{0},-\mathcal{T}/2 )~
  e^{-\int d^4 x~ \psi^\dagger (-D[U]) \psi} \cr
  &=
    \frac{1}{\mathcal{Z}_{\mathrm{eff}}[0]} \int\mathcal{D}
    U~e^{-S_{\mathrm{eff}}[U]}~\Gamma_{(TT_3)(SS_3)}^{\{f\}}
    \Gamma_{(TT_3)(SS_3)}^{\{g\}*} ~\prod_{i=1}^{N_{c}}G_{f_{i}g_{i}}(\bm{0}, 
    +\mathcal{T}/2|\bm{0},-\mathcal{T}/2) .
\label{eq:21}
\end{align}
$G_{f_{i}g_{i}}$ denotes the quark propagators in the background pion
field as 
\begin{align}
  G_{f_{i}g_{i}}(\bm{0}, +\mathcal{T}/2|\bm{0},-\mathcal{T}/2) &=
\langle \bm{0}, +\mathcal{T}/2, f_i |
  D[U]^{-1}i\gamma_{4}| \bm{0}, -\mathcal{T}/2, g_i\rangle  
= \int \frac{d\omega}{2\pi}~ 
\left\langle \bm{0}, f_i \left| \frac1{-i\omega+H} \right|
  \bm{0},g_i\right \rangle  e^{-i\omega \mathcal{T}}, 
\label{eq:22}
\end{align}
where $H$ denotes the single-particle Dirac Hamiltonian defined as   
\begin{align}
    H &=\gamma_{4}\gamma_{k}\partial_{k} +
             \buildrel\leftarrow\over F
             \gamma_{4}M_{0}U^{\gamma_{5}}(\bm{r})
             \buildrel\rightarrow\over  F 
\label{eq:23}
\end{align}
with the corresponding energy eigenvalue $E_n$ of the quark level
$|n\rangle$, i.e., $H|n\rangle = E_n(\omega)|n\rangle$. Note that the
eigenenergies are given as functions of $\omega$, which makes it
difficult to carry out the integral over $\omega$ analytically.

We will explicitly show that the valence-quark energy,
$E_{\mathrm{val}}$ is given by the product of the $N_c$ quark
propagators in the time-independent background field
$U_{\mathrm{cl}}(\bm{r})$, whereas the sea-quark
energy $E_{\mathrm{sea}}$ originates from the fermionic determinant
given in Eq.~\eqref{eq:11} with the time-independent $U$ field. 
We introduce the hedgehog symmetry for $\pi^a$~\cite{Pauli:1942kwa,
  Adkins:1983ya, Diakonov:1997sj}, $\bm{\pi} (\bm{r})=
\hat{\bm{n}}\Theta(r)$, where $\hat{\bm{n}}=\bm{r}/r$. It is a minimal
generalization of spherical symmetry. Then, we can express the chiral
field as   
\begin{align}
     U(r)= \exp\left[i \bm{\tau}\cdot
  \bm{\pi} (\bm{r})\right] = \exp\left[i (\bm{\tau}\cdot
  \hat{\bm{n}}) \Theta(r) \right] =  \cos\Theta(r) +
  i(\bm{\tau}\cdot\hat{\bm{n}})~\sin\Theta(r), 
\label{eq:24}
\end{align}
where $\Theta(r)$ is the profile function of the pion mean
field. Because of the hedgehog symmetry, the Hamiltonian commutes
neither with the isospin operator $\bm{T}$ nor with the total angular
momentum $\bm{J} = \bm{L}+ \bm{S}$. Instead, it commutes with the
grand spin $\bm{K}=\bm{J}+\bm{T}$. Thus, $K^2$ and $K_3$ are good
quantum numbers for the eigenenergies and eigenfunctions for the
Hamiltonian. The $N_c$ valence quarks fill the lowest state with
$K^P=0^+$. 

Had $E_n$ been independent of $\omega$, we would have expressed the
propagator as the spectral representation 
\begin{align}
  G_{f_{i}g_{i}}(\bm{0}, +\mathcal{T}/2|\bm{0},-\mathcal{T}/2) & =
 \Theta(\mathcal{T}) \sum_{E_n>0} e^{-E_n \mathcal{T}}
\psi_{nf_i} (\bm{0}) \psi_{ng_i}^\dagger (\bm{0}) 
 -\Theta(-\mathcal{T}) \sum_{E_n<0} e^{+E_n \mathcal{T}}
\psi_{nf_i} (\bm{0}) \psi_{ng_i}^\dagger (\bm{0})   
\label{eq:25}
\end{align}
as was obtained in Refs.~\cite{Diakonov:1987ty, Christov:1995vm}.
In the presence of the momentum-dependent dynamical quark mass,
however, we can only compute the integral numerically, using the 
residue theorem: 
\begin{align}
 G_{f_{i}g_{i}}(\bm{0}, +\mathcal{T}/2|\bm{0},-\mathcal{T}/2)
  &=-\int\frac{d\omega}{2\pi i} \sum_{n}\frac{\psi_{n,f_i}(\bm{0}) 
  \psi^{\dagger}_{n,g_i}(\bm{0})}{\omega+iE_{n}(\omega)
    }e^{-i\omega\mathcal{T}}\cr 
&=-\sum_{n}^{E_{n}>0}\Big(1+i\frac{\partial
    E_{n}(\omega)}{\partial \omega} \Big 
\vert_{\omega=-iE_{n}}\Big)^{-1}\psi_{n,f_i}(\bm{0})
\psi^{\dagger}_{n,g_i}(\bm{0}) e^{-E_{n}\mathcal{T}} \cr 
  &\underset{\mathcal{T} \rightarrow \infty}{\sim}
    z_{\mathrm{val}}~\psi_{{\rm{val}},f_i}(\bm{0})
\psi^{\dagger}_{{\rm{val}},g_i}(\bm{0})~ e^{-E_{\rm val} \mathcal{T}},  
\label{eq:26}
\end{align}
Note that the integration over the quark energy $\omega$ picks up
the discrete quark level. Taking the limit of the infinite Euclidean
time, i.e., $\mathcal{T}\to \infty$, we see that the quark propagator yields the 
valence-quark energy with the wavefunction for the valence quark
$\psi_{\rm val}(\bm{0})=\langle \bm{0}| \mathrm{val}\rangle$. The
prefactor $z_{\mathrm{val}}$, which arises from the residue theorem,
is expressed as  
\begin{align}
     z_{\mathrm{val}} &=\Big(1+i\frac{\partial E_{\rm val}(\omega)}{\partial
 \omega} \Big \vert_{\omega=-iE_{\rm val}}\Big)^{-1}. 
\label{eq:27}
\end{align}
Note that the value of $z_{\mathrm{val}}$ is smaller than one, i.e., 
$z_{\mathrm{val}} < 1$. This can be regarded as a ``wave-function
renormalization'' constant of the valence quark dressed by the pion
mean field. As will be shown later, $z_{\mathrm{val}}$ plays a
critical role to produce the baryon number properly.

Thus, the product of the $N_c$ quark propagators is
obtained as 
\begin{align} 
 \prod_{i=1}^{N_{c}}G_{f_{i}g_{i}}(\bm{0}, +\mathcal{T}/2|\bm{0},
  -\mathcal{T}/2) & 
   \underset{\mathcal{T} \rightarrow \infty}{\sim}
    ~\left(z_{\mathrm{val}}\psi_{\rm val}(\bm{0})\psi^{\dagger}_{\rm val}
    (\bm{0})\right)^{N_{c}}e^{-N_{c}E_{\mathrm{val}} \mathcal{T}},
\label{eq:28}
\end{align}
where the lowest positive energy state, which is the valence quark
state $\psi_{\rm val}$, only survives in the last step.  

The Dirac-continuum (sea-quark) energy is derived as  
\begin{align}
S_{\mathrm{eff}}[U]-S_{\mathrm{eff}}[0]
  &=-N_{c}\mathrm{Tr}\left[\log(D) - \log(D_{0}) \right]
    =-N_{c}\int\frac{d\omega}{2\pi}~{\mathrm{Tr}}
                       \left[\log(\omega+i H(\omega)) -
                       \log(\omega+i H_{0}(\omega))
                       \right]\cr  
  &=N_{c} \mathcal{T}\int\frac{d\omega}{2\pi} ~ \omega \sum_{n} 
    \left[\frac{1+i(\partial E_{n}(\omega)/\partial
    \omega)}{\omega+iE_{n}(\omega)} -\frac{
    1+i(\partial E_{n}^{(0)}(\omega)/\partial \omega)
    }{\omega+i E_{n}^{(0)}(\omega)}
    \right]=E_{\mathrm{sea}} \mathcal{T}, 
\label{eq:29}
\end{align}
where $S_{\mathrm{eff}}[0]$ is the effective action with $U\equiv 1$,
which comes from the normalization factor
$1/\mathcal{Z}_{\mathrm{eff}}[0]$ in Eq.~\eqref{eq:21}. The free
one-body Dirac operator is defined 
as $D_0 = i\slashed{\partial} + i M_{0}$. So, the free
Hamiltonian is defined as 
\begin{align}
H_0  = \gamma_{4}\gamma_{k}\partial_{k} + \gamma_{4}M_{0},
\label{eq:30}
\end{align}
which yields the free quark energies and levels, i.e.,
$H_{0}|n^{(0)}\rangle = E_n^{(0)}|n^{(0)}\rangle$. 

Therefore, in the limit of $\mathcal{T} \rightarrow\infty$, the
nucleon correlation function in Eq.~\eqref{eq:21} yields the classical
mass of 
the nucleon, $M_{\mathrm{cl}}$: 
\begin{align}
\Pi_{N}(\mathcal{T})\underset{\mathcal{T} \rightarrow \infty}{\sim}~ 
\int \mathcal{D}U~ e^{-N_{c}E_{\mathrm{val}} \mathcal{T}-E_{\mathrm{sea}}
\mathcal{T}} \sim  e^{-M_{\mathrm{cl}} \mathcal{T}},  
\label{eq:31}
\end{align}
which yields the normalization constant
$\mathcal{N}(\bm{p})=\sqrt{2M_N}$. Thus, the classical nucleon mass  
is correctly given by 
\begin{align}
    M_{\mathrm{cl}} &=\underset{\pi(x)}{\mathrm{min}} \left[N_c
                      E_{\mathrm{val}} + E_{\mathrm{sea}} \right],
\label{eq:32}
\end{align}
which indicates that the functional integration over $\pi^a$ is
carried out by the saddle-point approximation. 
The minimizing procedure for $M_{\mathrm{cl}}$ is performed in the
following steps: given a trial profile $\Theta(r)$, we diagonalize the
Dirac Hamiltonian in Eq.~\eqref{eq:19} to evaluate the eigenvalues and
eigenfunctions of the quarks. Then, inserting them into the classical
equation of motion, we derive a new profile function  
\begin{align}
    \frac{\partial S_{\rm eff}}{\partial \cos\Theta(r)} & =
    iN_{c}\int_{-\infty}^{\infty}\frac{d\omega}{2\pi}~\frac{\buildrel
    \leftarrow\over  F(\gamma_{4})\buildrel\rightarrow\over
     F}{\omega+i  H}-2\int d^{3}r~ S(r)=0, \cr  
    \frac{\partial S_{\mathrm{eff}}}{\partial\sin\Theta(r)}
  &=  iN_{c}\int_{-\infty}^{\infty}\frac{d\omega}{2\pi}~
    \frac{\buildrel \leftarrow\over F(i\gamma_{4}\gamma_{5}
    \bm{\tau}\cdot\hat{\bm{n}} )\buildrel\rightarrow\over F}{\omega+i
    H}   -2\int d^{3}r~P(r)=0, 
\label{eq:33}
\end{align}
where $S(r)$ and $P(r)$ are defined by 
\begin{align}
S(r) &:=  \cos\Theta(r) =-\frac{N_{c}}{2}z_{p}\langle {\rm val} \vert
\buildrel\leftarrow\over F(\gamma_{4})\buildrel
  \rightarrow\over F\vert {\rm val}\rangle
  +i\frac{N_{c}}{2}\int_{-\infty}^{\infty}
  \frac{d\omega}{2\pi}\sum_{n}\frac{\langle
  n\vert\buildrel \leftarrow\over F(\gamma_{4})
  \buildrel\rightarrow\over F\vert
  n\rangle}{\omega+i E_{n}}, \cr
P(r)&:= \sin\Theta(r) =-\frac{N_{c}}{2}z_{p}\langle {\rm val}
  \vert\buildrel\leftarrow\over
  F(i\gamma_{4}\gamma_{5} \bm{\tau}\cdot\hat{\bm{n}})
  \buildrel\rightarrow\over F\vert {\rm
  val}\rangle
  +i\frac{N_{c}}{2}\int_{-\infty}^{\infty}
  \frac{d\omega}{2\pi}\sum_{n}\frac{\langle
  n\vert  \buildrel\leftarrow\over
  F (i\gamma_{4}\gamma_{5}\bm{\tau}\cdot
  \hat{\bm{n}})\buildrel\rightarrow
  \over F\vert n\rangle}{\omega+i E_{n}}.
\label{eq:34}
\end{align}
We repeat this minimization procedure until we reach the minimum of
the classical mass. This is a well-known Hartree approximation. Thus,
the final profile function is obtained to be the pion mean-field solution. 

\begin{figure}[htp]
\centering
\centering
\includegraphics[width=0.45\columnwidth]{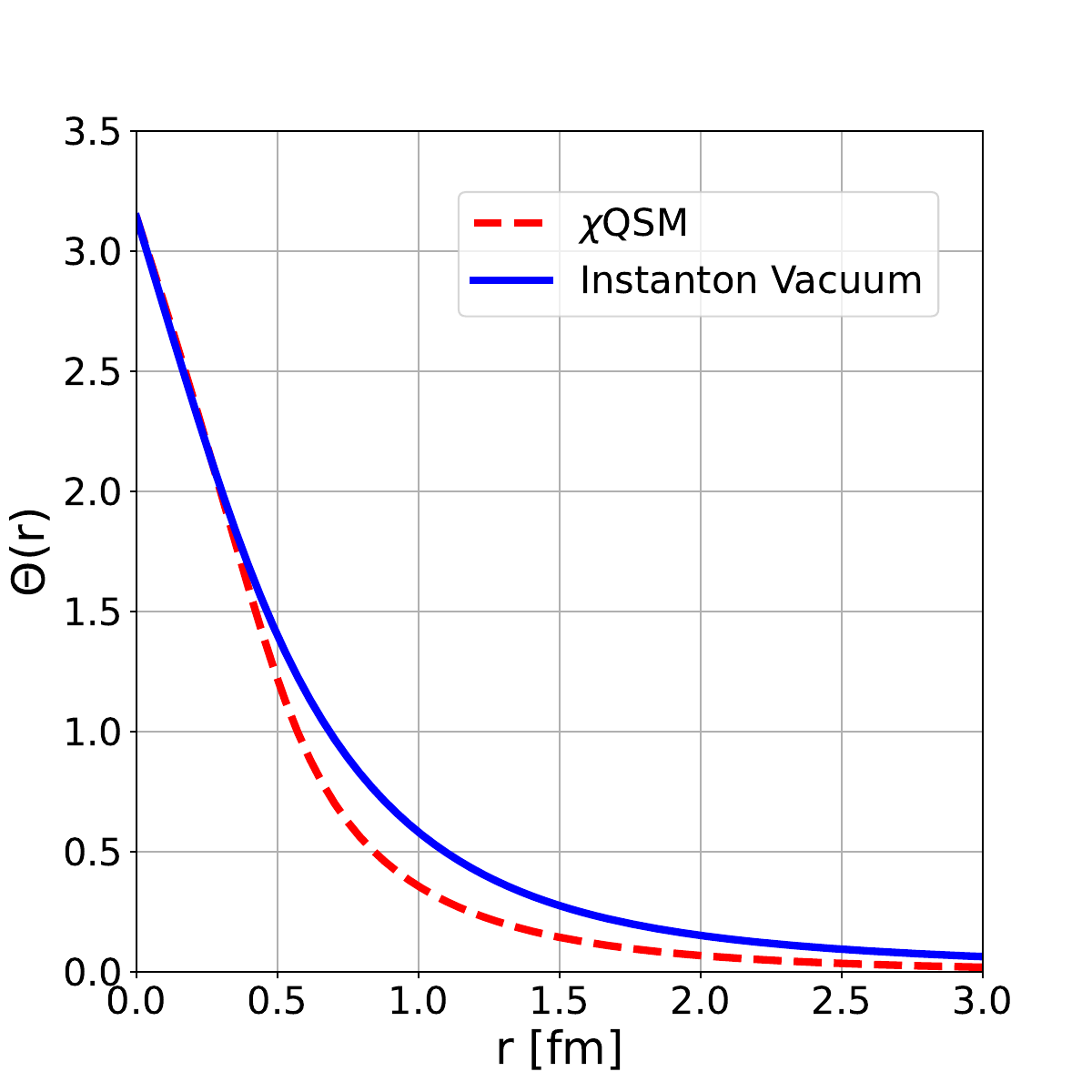}\;\;\;
\includegraphics[width=0.45\columnwidth]{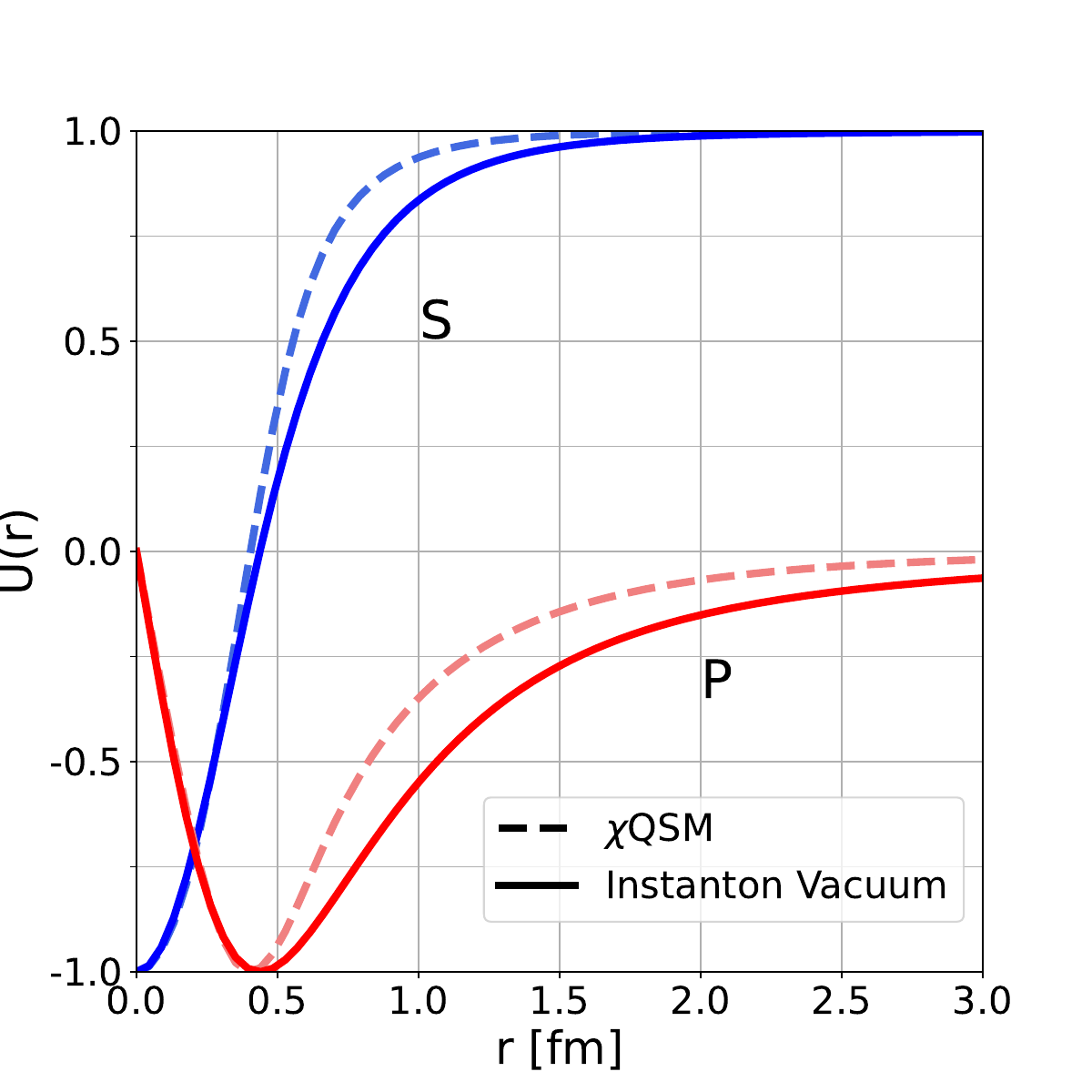}
\caption{\label{fig:3}%
The self-consistent profile function $\Theta(r)$ is drawn in
the left panel, whereas $S(r)=\cos[\Theta(r)]$ and
$P(r)=-\sin[\Theta(r)]$ are depicted in the right panel. The results
are compared with those from the $\chi$QSM~\cite{Christov:1995vm} with
$M_0=359$ MeV used.} 
\end{figure}
In Fig.~\ref{fig:3}, we draw the results for the self-consistent
profile function in the left panel, and $S(r)$ and $P(r)$ in the right
panel, which are defined in Eq.~\eqref{eq:34}. We compare the current
results with those from the local $\chi$QSM~\cite{Christov:1995vm}
with the same value of the dynamical quark mass at the zero virtuality
with $M_0=359$ MeV used. The present result for $\Theta(r)$ is broader
than that from the $\chi$QSM. Consequently, the numerical results for
$S(r)$ and $P(r)$ are also broader than those from
the local $\chi$QSM. Even though the profile function
looks different, the valence- and sea-quark energies are similar to 
each other. 

\begin{table}[h!]
\caption{\label{tab:1}%
The classical nucleon mass $M_{\mathrm{cl}}$.}
\centering
\renewcommand{\arraystretch}{2}
\setlength{\tabcolsep}{6pt}
\begin{tabular}{ccccc} \hline\hline
& $E_{\mathrm{val}}~{\rm [MeV]}$  & $N_c E_{\mathrm{val}}~\mathrm{[MeV]}$ &
  $E_{\mathrm{sea}}~\mathrm{[MeV]}$ & $M_{\mathrm{cl}}~\mathrm{[MeV]}$ \\ 
\hline
Present work & $257.4$ & $772.2$ & $495.8$ & $1268$ \\
$\chi$QSM ($M_0=359$ MeV)  & $240.8$ & $722.3$ & $532.9$ & $1255$  \\ 
$\chi$QSM ($M_0=420$ MeV)  & $203.5$ & $610.6$ & $645.3$ & $1256$  \\ 
\hline\hline
\end{tabular}
\end{table}
In Table~\ref{tab:1}, we list the numerical results for the valence-
and sea-quark energies, and the classical mass of the nucleon. 
The results are comparable to those from the $\chi$QSM with the same
value of the dynamical quark mass~\cite{Christov:1995vm} $M_0=359$
MeV. Note that $M_0$ is the free parameter in the $\chi$QSM, which was 
fixed by reproducing the mass spectrum of the baryons. The best value
for $M_0$ is known to be $420$ MeV. In the current theoretical
framework, however, $M_0$ is fixed by the gap equation from the
instanton vacuum, once the average size and the interdistance between
instantons are determined to be $\bar{\rho} =1/3$ fm and $\bar{R}
=0.98$ fm, respectively. 

It is interesting to compare the current results with those from
Ref.~\cite{Broniowski:2001cx}, which is technically similar to our
work. In Ref.~\cite{Broniowski:2001cx}, two different regulators were
employed, i.e., the Gaussian and monopole-type form factors with $M_0$
used as a free parameter. The cutoff mass was fixed by reproducing the
pion decay constant. $M_0=350$ MeV and the corresponding cutoff mass 
$\Lambda=627$ MeV were employed to get $E_{\mathrm{val}}=285$ MeV
(Gaussian) and $275$ MeV (monopole-type), which are comparable to the   
present value $E_{\mathrm{val}}=257.4$ MeV. The mass of the classical
nucleon from Ref.~\cite{Broniowski:2001cx} is also comparable to the
present one. 

In the calculation of the non-linear equation in Eq.~\eqref{eq:19},
$k^{2}$ may become negative,
as $\omega^{2}=-E^{2}_{\mathrm{val}}<0$. Although the quark form
factor $F(k)$ in Eq.~\eqref{eq:5} is derived from the 
instanton vacuum in Euclidean space, it can be analytically continued
into the $(k^{2}<0)$. Due to the branch point of the modified Bessel
functions $K_{0,1}(k=0)$ in Eq.~\eqref{eq:5}, $F(k)$ for $(k^{2}<0)$
can become complex, as 
\begin{align}
    {\rm Re}[F(k)]&=\frac{\pi}{2} z \frac{d}{dz} \big[I_{0}(z)Y_{0}(-i
                    z)+i I_{1}(z)Y_{1}(-i
                    z)\big]\Big{\vert}_{z=k\bar{\rho}/2}, \cr 
                    {\rm Im}[F(k)]&=\pi \big[2 z I_{0}(z)I_{1}(z)-(I_{1}(z))^{2}\big]
                     \Big{\vert}_{z=k\bar{\rho}/2}. 
\label{eq:35}
\end{align}
Although the momentum-dependent quark mass, $M(k)=M_{0}F^{2}(k)$ in
Eq.~\eqref{eq:4}, is not a physical quantity, we expect it to increase
monotonically in the $k^{2}<0$ region and to exhibit the behavior of a
smooth function. Therefore, we assume that the absolute value of the
quark form factor, $\vert F(k)\vert$, is used to determine the wave
function. As already pointed out by Diakonov et
al~\cite{Diakonov:1979nj}, the momentum-dependent dynamical form 
factor does not have a singularity on the mass shell of the quark,
even though the instantons cannot explain the quark confinement.

\section{Baryon number of the nucleon} \label{sec:4}
In the previous section, we showed that the baryon
number $B=1$ is not spoiled by the saddle-point approximation. 
In the current framework, the baryon number of the nucleon is
determined by the $N_c$ valence quarks~\cite{Diakonov:1987ty,
  Christov:1995vm}, which means that each valence quark carries baryon
number $1/N_c$. This is distinguished from the skyrme model, where the
baryon number is identified as a topological winding number related to
the Wess-Zumino-Witten term that generates the topological current.

We will show that the baryon number of the nucleon is solely
determined by the $N_c$ valence quarks, computing the three-point
correlation functions for the temporal component of the 
vector current. However, it is well known that gauge
invariance is broken in the presence of a nonlocal
interaction~\cite{chretien:1954we}, which results in the
non-conservation of the vector and axial-vector currents. So, the
Ward-Takahashi identity is not satisfied. If we ignore this problem,
we would have incomplete expressions for any matrix elements for the
vector or axial-vector current\footnote{For example, the expression
  for the pion decay constant becomes
  incomplete~\cite{Pagels:1979hd}.}. To restore the gauge invariance,
a path-dependent gauge connection was introduced in the nonlocal
interactions~\cite{Holdom:1989jb, Holdom:1992fn, Bos:1991sz,
  Bowler:1994ir, Broniowski:1999bz}. Within the current theoretical  
framework, this problem was already pointed out in 
Ref.~\cite{Pobylitsa:1989uq}. The reason
arises from an incomplete summation of higher-order terms with respect
to the density of the instanton medium when the correlation   
functions were evaluated from the instanton
vacuum~\cite{Diakonov:1985fjw}. It brought about the non-conservation
of the vector and axial-vector currents~\cite{Pobylitsa:1989uq}. 
In Ref.~\cite{Musakhanov:2002xa}, the gauge connection was introduced
from the very beginning in the derivation of the effective partition 
function. It was shown that the ordinary derivative in it 
can be replaced with the covariant one when the strength of a current
is weak. Thus, the gauged effective low-energy partition function can
be expressed as   
\begin{align}
  Z_{\mathrm{eff}}[U,\mathcal{V}] =  \int \mathcal{D}U
  ~\exp(-S_{\mathrm{eff}}[U, \mathcal{V}]) 
\label{eq:36}
\end{align}
with the gauged NE$\chi$A
\begin{align}
S_{\mathrm{eff}}[U,\mathcal{V}] =-N_c\mathrm{Tr}\log \left[
  i\slashed{\nabla} +   i M_0  \overleftarrow{F}(i\nabla)
  U^{\gamma_5}[\pi^a(x)]   \overrightarrow{F}(i\nabla)   \right].
\label{eq:37}
\end{align}
Here, $\nabla$ denotes the covariant derivative written as 
\begin{align}
\nabla_{\mu}=\partial_{\mu}-i B_\mu,  
\label{eq:38}
\end{align}
where $B_\mu$ is the baryon current defined as $B_\mu:={N_c}^{-1}
\mathcal{V}_{\mu}$. Thus, the gauged partition function is nothing but
the original one with the gauged form factor $F(i\partial_{\mu} + 
N_c^{-1} \mathcal{V}_\mu)$. Interestingly, this is just the same as
the minimal substitution. 

We introduce a generating functional (in Euclidean space) for the
baryon number defined by 
\begin{align}
{\mathcal{W}}[\eta ^{\dagger },\eta ,\mathcal{V}_{4}] 
=~&  \frac{-i}{\mathcal{Z}_{\mathrm{eff}}[0]}\int  \mathcal{D}\psi ^{\dagger}\mathcal{D}\psi
\mathcal{D}U~ \exp \left[-\int d^{4}z~ \Big\{ \psi ^{\dagger } 
\Big(-i\slashed{\nabla}-iM_0 \overleftarrow{F}(i\nabla) 
U^{\gamma_{5}}(z) \overrightarrow{F}(i\nabla)\Big)\psi +i\eta^{\dagger}
\psi  +i\psi ^{\dagger }\eta \Big\} \right] \cr
=~&  \frac{1}{\mathcal{Z}_{\mathrm{eff}}[0]}\int \mathcal{D} U~ 
{\det} \left[i\slashed{\nabla}+iM_0 \overleftarrow{F}(i\nabla) 
U^{\gamma_{5}}(z) \overrightarrow{F}(i\nabla)\right]  \cr
&\times\exp \left[ -\int d^{4}x d^{4}y~ 
\eta ^{\dagger }_{\alpha }(y) \left\langle y,\alpha\left| 
\Big(-\slashed{\nabla}-M_0 \overleftarrow{F}(i\nabla) 
U^{\gamma_{5}}(z) \overrightarrow{F}(i\nabla)\Big)^{-1} 
\right| x,\beta \right\rangle\eta_{\beta }(x)\right].   
\label{eq:39}
\end{align}
Then, the three-point correlation function $\mathcal{K}$ is evaluated by 
the functional derivatives with respect to the source
fields $\eta^\dagger$, $\eta$, and $\mathcal{V}_4$: 
\begin{align}
\mathcal{K} 
=&~  \Gamma_{(TT_{3})(S S'_{3})}^{\{f\}}
\Gamma_{(TT_{3})(S S_{3})}^{\{g\}*}  \prod_{i=1}^{N_c}
\Big(\frac{\delta }{\delta \big(-i\eta ^{\dagger }_{ f_{i}}(\bm{y},y_{4})\big)}\Big) 
\frac{\delta }{\delta (-i\mathcal{V}_{4}(0))}
\left.  {\mathcal{W}}[\eta ^{\dagger },\eta ,\mathcal{V}_{4}]\prod_{i=1}^{N_c}
\Big(\frac{\stackrel{\leftarrow }{\delta }}{\delta  
\eta _{g_{i}}(\bm{x},x_{4})}\gamma_{4}\Big)\right|_{\eta^{\dagger },
\eta ,\mathcal{V}_{4}=0}  \cr
=&~ \frac{1}{\mathcal{Z}_{\mathrm{eff}}[0]} \int \mathcal{D}U~
\Bigg[ e^{-S_{\mathrm{eff}}} \Big\{\frac{\delta }{\delta (-i\mathcal{V}_{4}(0))}
\Big(\prod_{i=1}^{N_c} (-i)\left\langle y,f_{i}\left| \left(\partial_{t}+H(\mathcal{V}_{4})-iN_{c}^{-1}\mathcal{V}_{4}\right)^{-1} 
\right| x,g_{i} \right\rangle\Big)\Big\}\cr
&\qquad\qquad\qquad\quad +\frac{\delta }{\delta (-i\mathcal{V}_{4}(0))}
\Big(e^{-S_{\mathrm{eff}}}\Big)\prod_{i=1}^{N_c} (-i)\left\langle y,f_{i}
\left| \left(\partial_{t}+H(\mathcal{V}_{4})-iN_{c}^{-1}\mathcal{V}_{4}\right)^{-1} 
\right| x,g_{i} \right\rangle\Bigg]\Bigg\vert_{\mathcal{V}_{4}=0}\cr
=&~ \frac{1}{\mathcal{Z}_{\mathrm{eff}}[0]} \int \mathcal{D}U~
\Gamma_{(TT_{3})(S S'_{3})}^{\{f\}}\Gamma_{(TT_{3})(S S_{3})}^{\{g\}*}
(-i)^{N_{c}}\prod ^{N_{c}}_{k=2} G_{f _{k}g _{k}}(y|x)
~e^{-S_{\mathrm{eff}}}\cr
& ~\times   \left[(-1)G_{f_1 f}  (y|0)\left\{\delta_{fg}+\left(
i\gamma_{4}\big(\buildrel\leftarrow \over F_{4}
M_{0}U^{\gamma_{5}}\buildrel\rightarrow\over
F+\buildrel\leftarrow\over FM_{0} U^{\gamma_{5}} \buildrel
\rightarrow \over F_{4}\big)\right)_{fg} \right\}G_{g g_{1}}(0|x)\right.\cr 
& \quad\quad +  \left. G_{f _{1}g _{1}}(y|x)
\mathrm{Tr} \left\{\left(1+i\gamma_{4}\big(\buildrel\leftarrow\over F_{4}
M_{0}U^{\gamma_{5}}\buildrel\rightarrow\over
F+\buildrel\leftarrow\over FM_{0} U^{\gamma_{5}} \buildrel
\rightarrow \over F_{4}\big)\right)  G(0|0) \right\} \right] ,
\label{eq:40}
\end{align} 
where $F_{4}$ is defined as the derivative of $F(k)$ with respect to
$k_4$, $F_4= \partial F/\partial k_4$. Note that the functional
derivative with respect to $\mathcal{V}_4$ acts also on the quark form
factor $F(\partial_\mu + i N_c^{-1} \mathcal{V}_{4})$, so that we additionally
have the term $i\gamma_{4}(\buildrel\leftarrow\over F_{4}
M_{0}U^{\gamma_{5}}\buildrel\rightarrow\over
F+\buildrel\leftarrow\over FM_{0} U^{\gamma_{5}} \buildrel
\rightarrow \over F_{4})$, which makes the Ward identity satistied. 
The first term in the bracket
of Eq.~\eqref{eq:40} is the valence-quark contribution to the
baryon number, whereas the second term represents the sea-quark
contribution that is expected to vanish.

Having carried out straightforward calculations of
Eq.~\eqref{eq:40}, we respectively obtain the expressions for
valence- and sea-quark contributions to the 
baryon number
\begin{align}
\langle \hat{B}\rangle_{N,\mathrm{val}} &=\Big\langle \psi^{\dagger}(0)\Big(-iN_{c}^{-1}\gamma_{4}+N_{c}^{-1}(\buildrel\leftarrow\over F_{4}
M_{0}U^{\gamma_{5}}\buildrel\rightarrow\over
F+\buildrel\leftarrow\over FM_{0} U^{\gamma_{5}} \buildrel
\rightarrow \over F_{4})\Big)\psi(0)\Big\rangle_{\rm val}\cr
& =
\oint_{c}\frac{d z}{2\pi i}~\frac{1+i\gamma_{4}(\buildrel\leftarrow\over F_{4} 
   M_{0}U^{\gamma_{5}}\buildrel\rightarrow\over
   F+\buildrel\leftarrow\over FM_{0} U^{\gamma_{5}} \buildrel
\rightarrow \over F_{4}) }{z +
  iE_{\mathrm{val}}(z)} = 
  \left( 1+i\frac{\partial E_{\rm val}(z)}{\partial z}\right)\bigg{\vert}_{z=-iE_{\rm val}}(z_{\mathrm{val}})=1,\cr \cr
\langle \hat{B}\rangle_{N,\mathrm{sea}} &=\Big\langle \psi^{\dagger}(0)\Big(-iN_{c}^{-1}\gamma_{4}+N_{c}^{-1}(\buildrel\leftarrow\over F_{4}
M_{0}U^{\gamma_{5}}\buildrel\rightarrow\over
F+\buildrel\leftarrow\over FM_{0} U^{\gamma_{5}} \buildrel
\rightarrow \over F_{4})\Big)\psi(0)\Big\rangle_{\rm sea}\cr
&=
\int_{-\infty}^{\infty} \frac{d \omega}{2\pi i}
   \left(\frac{1+i\gamma_{4}(\buildrel\leftarrow\over F_{4} 
   M_{0}U^{\gamma_{5}}\buildrel\rightarrow\over
   F+\buildrel\leftarrow\over FM_{0} U^{\gamma_{5}} \buildrel
\rightarrow \over F_{4}) }{\omega + iH}
   - \frac{1+i\gamma_{4}(\buildrel\leftarrow\over F_{4}
   M_{0}\buildrel\rightarrow\over
   F+\buildrel\leftarrow\over FM_{0} \buildrel
\rightarrow \over F_{4})}{\omega + iH_{0}}\right) \cr
& =\int_{-\infty}^{\infty}\frac{d\omega}{2\pi i}
                          \sum_{n}\frac{1+i(\partial E_{n}/\partial
                          \omega )}{\omega+ iE_{n}(\omega)}-
 \int_{-\infty}^{\infty}\frac{d\omega}{2\pi i}
                          \sum_{n}\frac{1+i(\partial
                          E_{n}^{(0)}(\omega) /\partial
                          \omega )}{\omega+ iE_{n}^{(0)}(\omega)}  =0.
\label{eq:41}
\end{align}
Therefore, the total baryon number is entirely determined by the
valence contribution. 

The present effective chiral theory of the nucleon interpolates
between the nonrelativistic quark model (NRQM) and the Skyrme model. 
This can be shown by introducing the dimensionless parameter $M_0R$,
where $R$ is the soliton size. 
If we turn off the pion mean field ($M_0R=0$), the $N_c$ valence
quarks lie in the state with the valence quark energy
$E_{\mathrm{val}}=M_0$. If we increase the value of $M_0R$, the
valence level filled with the $N_c$ valence quarks will get bound by the
pion mean field. If we further increase it, the level 
crosses zero energy and turns negative. When $M_0R\gg 1$,  
the $N_c$ valence quarks dive into the Dirac sea, and a topological
soliton emerges. This can be understood by the derivative expansion:
expanding the imaginary part of the effective chiral action in
Eq.~\eqref{eq:11} with respect to $\partial_k U/M_0$ to the order of
$\mathcal{O}(p^5)$, the WZW action arises as  
\begin{align}
\mathrm{Im} S_{\mathrm{eff}}[U] = N_c \Gamma = \frac{iN_c}{240\pi^2}
  \int d^4 x~ \varepsilon_{\alpha\beta\gamma\delta \rho}
  \mathrm{Tr}[L_\alpha L_\beta L_\gamma L_\delta L_\rho] +
  \mathcal{O}(p^6), 
\label{eq:42}
\end{align}
where $L_\alpha = U^\dagger \partial_\alpha U$.
Under the U(1) gauge transformation, the WZW term generates the baryon
number current as a N{\"o}ther current~\cite{Witten:1983tw,
  Witten:1983tx}. Thus, the baryon number is given by   
\begin{align}
\hat{B}(U) = -\frac1{24\pi^2} \int d^3 x~
  \varepsilon_{ijk}\mathrm{Tr}[L_iL_jL_k] = 1,   
\label{eq:43}
\end{align}
where $L_i = U^\dagger \partial_i U$. As pointed out by Diakonov and
Eides~\cite{Diakonov:1983bny}, the WZW term originates from the
fermionic determinant in QCD, which indicates the physical implications
of the derivative expansion of Eq.~\eqref{eq:12}. 

Concerning the WZW term, Ball and Ripka~\cite{Ball:1993ak} showed that
the WZW term is correctly obtained with the momentum-dependent quark
mass. On the other hand, the constant dynamical quark mass requires
introducing a regularization scheme, which spoils the WZW 
term. To avoid this inconsistent problem, the regularization was
only applied to the real part whereas the imaginary part is not
regularized. This means that the present framework does not suffer
from this inconsistency problem.     

\section{Zero-mode Quantization \label{sec:5}}
When we derived the mass of the classical nucleon in Sec.~\ref{sec:3},
we used the saddle-point approximation. It is a plausible
approximation, since mesonic quantum fluctuations, which are of the 
$1/N_c$ order, are suppressed in the large $N_c$ limit. On the other
hand, there are yet another fluctuations of the pion mean field in the
zero-mode directions, which are not at all small. Thus, we must
consider these quantum fluctuations completely. The zero modes are
related to continuous symmetries of the pion field, which are
translations, spatial and isospin rotations. Since the translational
zero modes give the linear momentum of the nucleon, it must be
considered when one computes form factors of the nucleon. The spatial
and isospin rotational zero modes assign the proper spin and isospin
quantum numbers to the classical nucleon~\cite{Adkins:1983ya,
  Jain:1984gp,Diakonov:1987ty, Diakonov:1997sj}.     

To assign the quantum number of the nucleon to the classical one 
that was obtained in the previous section, we consider the rotational
zero modes~\cite{Adkins:1983ya}. It does not affect the classical
nucleon mass $M_{\rm cl}$, since the hedgehog symmetry in
Eq.~\eqref{eq:24} makes the chiral field isotropic both in spatial and
flavor spaces. Thus, we consider time-dependent slow rotation of the
minimized chiral field $U_{\mathrm{cl}}(\bm{x})$ in both flavor and
ordinary spaces as 
\begin{align}
\tilde{U}(\bm{x}, t) = A(t) U_{\mathrm{cl}}(O\bm{x}) A^\dagger (t) =
A(t) S(t) U_{\mathrm{cl}}(\bm{x}) (A(t)  S(t))^{-1} = R(t)
U_{\mathrm{cl}}(\bm{x}) R^{-1} (t),
\label{eq:44} 
\end{align}
where $A$ is a flavor SU(2) unitary matrix that can be considered as 
left multiplication, and the SU(2) unitary matrix $S$ generates the
rotation in ordinary space, which can be right multiplication. Thus,
we can consider a single SU(2) rotation $R(t)$. The pion 
mean field remains invariant as far as the ordinary rotation is
compensated by the flavor rotation~\cite{Mazur:1984yf,
  Jain:1984gp,Diakonov:1987ty, Diakonov:1997sj}.

Even though we transform the Dirac operator in Eq.~\eqref{eq:13} by
replacing $U$ with Eq.~\eqref{eq:44}, the eigenvalues and
eigenfunctions of the quarks are not changed. So, we can write
Eq.~\eqref{eq:13} as    
\begin{align}
D[\tilde{U}] &= i\slashed{\partial} + 
i M_0\buildrel\leftarrow\over F(i\partial)
R(t)U^{\gamma_5}R^\dagger(t) \buildrel \rightarrow \over F(i\partial) 
\cr
& = R(t)\left[R^\dagger(t) i\slashed{\partial} R(t) + i M_0 R^\dagger(t)
\buildrel    \leftarrow \over F(i\partial) 
R(t)U^{\gamma_5} R^\dagger(t) \buildrel \rightarrow \over
F(i\partial) R(t)\right] R^\dagger(t).
\label{eq:45}
\end{align}
The first term in Eq.~\eqref{eq:45} can be expressed as 
\begin{align}
R^\dagger (t)  i\slashed{\partial} R(t) = i\slashed{\partial} +
  i\gamma_4 R^\dagger(t) \dot{R}(t)=i\slashed{\partial} -
  \gamma_4 \Omega,
  \label{eq:46}
\end{align}
where we define the right and left angular velocities respectively as  
\begin{align}
\Omega&= -i R^{\dagger} \dot{R} = \frac{1}{2}\Omega^{a}\sigma^{a},
\;\;\;
        \Omega_a = -i \mathrm{Tr}(R^\dagger \dot{R} \sigma^a),\cr
\tilde{\Omega}&= -i \dot{R} R^{\dagger} =
                \frac{1}{2}\tilde{\Omega}^{a}\tau^{a}, 
\;\;\;
        \tilde{\Omega}^a = -i \mathrm{Tr}(\dot{R} R^\dagger
                \tau^a),
\label{eq:47} 
\end{align}
where $\sigma^a$ and $\tau^a$ denote the Pauli matrices for the spin
and isospin.  

It is more involved to deal with the second term of Eq.~\eqref{eq:45}.
Expanding the form factor $ F(k)$ with respect to $k$ as 
\begin{align}
 F(i\partial) &= \sum_{n=1}^\infty  \frac{i^n}{n!}
                (\partial_{\mu_1}\cdots 
                \partial_{\mu_n}) \left.\frac{\partial^n 
                F(k)}{\partial k_{\mu_1}\cdots k_{\mu_n} }\right|_{k^2=0},
\label{eq:48}                
\end{align}
and acting it on $R$ to get the corrections to the order
$\mathcal{O}(\Omega^2)$, we express Eq.~\eqref{eq:45} as  
\begin{align}
  D[\tilde{U}] &=   R \left(i\partial_t + iH - \frac12 \Omega^a t^a 
                 -\frac14 \Omega^a \Omega^b  T^{ab} \right)R^\dagger,
\label{eq:49}                 
\end{align}
where $t^a$ and $T^{ab}$ are defined as
\begin{align}
  t^a & = \tau^a + \buildrel\leftarrow\over
        F_{4}\tau^{a} \gamma_{4}U^{\gamma_{5}}
        \buildrel\rightarrow\over F 
        + \buildrel\leftarrow\over F\gamma_{4}U^{\gamma_{5}}
       \tau^{a}\buildrel\rightarrow\over F_{4} \cr
        T^{ab}&= \tau^a\tau^b
                \buildrel\leftarrow\over F_{44} \gamma_{4} U^{\gamma_{5}}
                \buildrel \rightarrow\over F
                + 2 \buildrel\leftarrow\over 
         F_{4} \tau^{a}\gamma_{4}U^{\gamma_{5}}
         \tau^{b}\buildrel\rightarrow\over F_{4}
         + \buildrel\leftarrow\over F\gamma_{4}U^{\gamma_{5}}
                \tau^a \tau^b  \buildrel\rightarrow\over F_{44}  .
\label{eq:50}                
\end{align}
Here, $F_{44}\equiv\partial^2 F/\partial \omega^2$.

Thus, the quark propagator receives the $1/N_c$ rotational corrections
as follows:  
\begin{align}
 & G_{fg} (\bm{0}, \mathcal{T}/2|\bm{0}, -\mathcal{T}/2)[\tilde{U}] =
\langle \bm{0}, \mathcal{T}/2 | D[RU_{\mathrm{cl}}R^\dagger]^{-1} i\gamma_{4}
\ |\bm{0}, -\mathcal{T}/2\rangle_{fg} \cr
&= \left[
R(\mathcal{T}/2) \left\langle \bm{0}, \mathcal{T}/2\left| \frac1{\partial_t + H + i
    \Omega^a t^a/2  +i \Omega^a \Omega^b  T^{ab}/4} \right|\bm{0},
-\mathcal{T}/2 \right\rangle R^\dagger(-\mathcal{T}/2)\right]_{fg}  \cr                                            
&\simeq -\Big[R(\mathcal{T}/2) z_{\mathrm{val}} \psi_{\mathrm{val},f}(\bm{0})
\psi^{\dagger}_{\mathrm{val},g} (\bm{0}) 
R^{\dagger}(-\mathcal{T}/2)\Big]e^{-E_{\mathrm{val}} \mathcal{T}} 
\left(1-\frac12 I_{\mathrm{val}} \delta^{ab}                                  
\int_{-\mathcal{T}/2}^{\mathcal{T}/2} dt~ \Omega^a(t)
\Omega^b(t) \right) ,
\label{eq:51}                                              
\end{align}
where $I_{\mathrm{val}}$ is the valence-quark contribution to the
moment of inertia for the pion mean field, which is
expressed as 
\begin{align}
I_{\mathrm{val}}\delta^{ab} &= 
\frac{N_{c}}{2}\sum_{n}\frac{z_{\mathrm{val}}}{E_{n}-E_{\mathrm{val}}}
\langle \mathrm{val}\vert t^{a} \vert n\rangle \langle n
\vert t^{b}\vert \mathrm{val}\rangle\big{\vert}_{\omega=-i
E_{\mathrm{val}}}-\frac{N_{c}}{8} z_{\mathrm{val}}\langle
\mathrm{val}  \vert T^{ab} \vert \mathrm{val}
\rangle \big{\vert}_{\omega=-i E_{\mathrm{val}}}  .
\label{eq:52}                       
\end{align}
The effective chiral action under flavor rotation is expressed as
\begin{align}
S_{\mathrm{eff}}[\tilde{U}] & = E_{\mathrm{sea}} \mathcal{T} -
N_c\mathrm{Tr}\left[\log\left(i\partial_t + iH  - \frac12 \Omega^a t^a   
-\frac14 \Omega^a \Omega^b  T^{ab} \right) -\log\left(i\partial_t
+ iH\right)\right] \cr
&= E_{\mathrm{sea}} \mathcal{T} + \frac12 I_{\mathrm{sea}} \delta^{ab}
\int_{-\mathcal{T}/2}^{\mathcal{T}/2}dt~ \Omega^a (t) \Omega^b(t),
\label{eq:53}                                                            
\end{align}
where $I_{\mathrm{sea}}$ is the sea-quark contribution to the
moment of inertia 
\begin{align}
I_{\mathrm{sea}}\delta^{ab} &= \frac{N_{c}}{4} \sum_{n,m}^{n\neq m}
\int \frac{d\omega}{2\pi}~  \frac{\langle m\vert t^{a}\vert
n\rangle}{\omega+i E_{n}}  \frac{\langle n\vert t^{b}\vert
m\rangle}{\omega+i E_{m}} - \frac{N_{c}}{8}
\sum_{n}  \int\frac{d\omega}{2\pi i}~
\frac{\langle n\vert T^{ab}\vert n\rangle}{
\omega+i E_{n}}.
\label{eq:54}
\end{align}
Thus, the total moment of inertia for the nucleon is obtained as 
\begin{align}
    I=I_{\rm val}+I_{\rm sea}.
\label{eq:55}
\end{align}
If we switch off the momentum dependence of the dynamical quark mass
$F(k)=1$, then the expressions for the moments of inertia are reduced
to those in local effective chiral theory of the
nucleon~\cite{Diakonov:1987ty} 
\begin{align}
I_{\mathrm{val}}&=\frac{N_{c}}{6}\sum_{n} 
\frac{\left|\langle \mathrm{val}| \bm{\tau}| n\rangle\right|^2}
{E_{n}-E_{\mathrm{val}}},\;\;\;
I_{\mathrm{sea}}=\frac{N_{c}}{6}\sum_{n,m}
\frac{\left|\langle m| \bm{\tau}| n\rangle\right|^2}{E_n-E_m}
\Theta(E_n) \Theta(-E_m).
\label{eq:56}
\end{align}

Collecting all the expressions given above, we can express the
zero-mode quantized nucleon correlation function as
\begin{align}
  \Pi_N(\mathcal{T}) &= \Gamma ^{\{f \}}_{(TT'_{3})(SS_3)}
       \Gamma ^{\{g \}*}_{(TT_{3})(SS_3)}
        \int \mathcal{D}R(t) ~\prod_{i=1}^{N_c}~ [R(\mathcal{T}/2)\psi_{\mathrm{val},f_i} (\bm{0}) 
        \psi_{\mathrm{val},g_i}^\dagger (\bm{0}) R^\dagger (-\mathcal{T}/2)] \cr
  &\quad\times
   \exp\left[ -(N_c E_{\mathrm{val}}
               +E_{\mathrm{sea}} )\mathcal{T} - \frac{I}{2}\sum_{a=1}^3 \int_{-\mathcal{T}/2}^{\mathcal{T}/2}
               dt~\Omega_a^2(t)    \right].
\label{eq:57}
\end{align}
The second term in the collective action given in Eq.~\eqref{eq:53}
is nothing but the Lagrangian of the spherical top~\cite{Landau:QM}   
\begin{align}
L_{\mathrm{rot}} &= \frac{I}{2} \sum_{a=1}^3 \Omega_a^2 = \frac{I}{2}
\sum_{a=1}^3 \tilde{\Omega}_a^2 , 
\label{eq:58}
\end{align}
where $\Omega_a$ and $\tilde{\Omega}^a$ are given in
Eq.~\eqref{eq:47}. 
The corresponding Hamiltonian is then expressed as 
\begin{align}
H_{\mathrm{rot}} &= \bm{\Omega} \cdot \bm{S} - L_{\mathrm{rot}} =
\frac1{2I} (S_a)^2 =  \tilde{\bm{\Omega}} \cdot \bm{T} - 
L_{\mathrm{rot}} =  \frac1{2I}  (T^A)^2
\label{eq:59}
\end{align}
with two sets of angular momenta conjugate to $\Omega_a$ and
$\tilde{\Omega}_A$ 
\begin{align}
  S_a &= \frac{\partial L_{\mathrm{rot}}}{\partial \Omega_a},
  \;\;\;
    T_A = \frac{\partial L_{\mathrm{rot}}}{\partial \tilde{\Omega}_A}.
\label{eq:60}
\end{align}

It is now straightforward to quantize the rotational Hamiltonian. 
Two sets of the angular momenta can be identified as the spin
operator $S_i$ and the isospin operator $T_A$, respectively, since
they can be considered as a right and left multiplication operators
respectively, as discussed previously, when the spin and isospin
operators act on $R$. They satisfy the following  
transformation rules 
\begin{align}
e^{i\bm{S}\cdot \bm{\theta}} R e^{-i\bm{S}\cdot \bm{\theta}} &= 
R e^{i\bm{\sigma}\cdot  \bm{\theta}/2} ,\cr
e^{i\bm{T}\cdot \bm{\theta}} R e^{-i\bm{T}\cdot \bm{\theta}} &= 
 e^{-i\bm{\tau}\cdot  \bm{\theta}/2} R,
\label{eq:61}
\end{align}
where $S_a$ and $T_A$ satisfy the following commutation
rules~\cite{Diakonov:1987ty, Diakonov:1997sj} 
\begin{align}
[S_a, ~S_b] &= i\varepsilon_{abc}S_c,\;\;\;
[T_A,~T_B] = i  \varepsilon_{ABC}T_C,\;\;\;
[S_a, T_B] =0.
\label{eq:62}
\end{align}
Thus, the quantized rotational Hamiltonian can be expressed as
\begin{align}
H_{\mathrm{rot}} &= \frac{S_a^2}{2I}  = \frac{T_A^2}{2I},
\label{eq:63}
\end{align}
which can also be called as the collective Hamiltonian.  

The normalized simultaneous collective eigenfunction of
$H_{\mathrm{coll}}$ is written in terms of the SU(2) Wigner $D$
funcion
\begin{align}
  \Psi_{(TT_3)(SS_3)} (R) &= \sqrt{2S+1}  (-1)^{T+T_3}
                        D_{-T_3 S_3}^{S=T}(R),
\label{eq:64}
\end{align}
and the corresponding eigenvalues are given as
\begin{align}
E_{\mathrm{rot}} &= \frac{S(S+1)}{2I} = \frac{T(T+1)}{2I},
\label{eq:65}
\end{align}
which can be called as the rotational $1/N_c$ corrections to
$M_{\mathrm{cl}}$, since $I$ is proportional to $N_c$.  

It is natural that the collective wave functions can also be written as   
\begin{align}
\Psi_{(TT_3)(SS'_3)}^*(R(\mathcal{T}/2)) &= \Gamma ^{\{f \}}_{(TT_{3})(SS'_3)}
  \prod_{i=1}^{N_c} (R(\mathcal{T}/2)\psi_{\mathrm{val}}
  (\bm{0}) )_{f_i} ,\cr
\Psi_{(TT_3)(SS_3)}(R(-\mathcal{T}/2)) &= \Gamma ^{\{g *\}}_{(TT_{3})(SS_3)}
 \prod_{i=1}^{N_c}   (\psi_{\mathrm{val}}^\dagger (\bm{0}) 
R^\dagger (-\mathcal{T}/2))_{g_i} ,
\label{eq:66}                          
\end{align}
since $\Gamma ^{\{g *\}}_{(TT_{3})(SS_3)} $ and $\Gamma ^{\{f
  \}}_{(TT_{3})(SS'_3)}$ project the $N_c$ valence-quark states to
the corresponding baryon states with definite quantum numbers given.
Noting that the matrix element of the evolution operator can be
expressed as
\begin{align}
&\int_{R(-\mathcal{T}/2)}^{R(\mathcal{T}/2)} \mathcal{D}R(t)~
                \exp\left[ - \frac{1}{2I} 
  \int_{-\mathcal{T}/2}^{\mathcal{T}/2}  dt~ S_a^2 
  \right]  = \left\langle R(\mathcal{T}/2) \left |  \exp\left[-
             \frac{1}{2I} S_a^2 \right] \right |R(-\mathcal{T}/2)
             \right \rangle \cr
&= \sum_{SS_3T_3} \psi_{(TT_3)(SS_3)}(R(\mathcal{T}/2))
             \psi_{(TT_3)(SS_3)}^*(R(-\mathcal{T}/2)) \exp\left[-
             \frac{1}{2I} S(S+1) \right], 
\label{eq:67}
\end{align}
and imposing the boundary conditions at $t=-\mathcal{T}/2$ and
$t=\mathcal{T}/2$,  we can express Eq.~\eqref{eq:57} as 
\begin{align}
  \Pi_N(\mathcal{T}) &= \int \mathcal{D}R(t)~
                       \Psi_{(TT_3)(SS'_3)}^*(R_1)
\Psi_{(TT_3)(SS_3)}(R_2) ~\exp\left[ -M_{\mathrm{cl}} \mathcal{T}
- \frac{1}{2I} \int_{-\mathcal{T}/2}^{\mathcal{T}/2}  dt~S_a^2    \right] \cr
  &=  \int dR_1 dR_2~ \Psi_{(TT_3)(SS'_3)}^*(R_1) \Psi_{(TT_3)(SS_3)}(R_2) 
    \Psi_{(TT_3)(SS_3)}(R_1) \Psi_{(TT_3)(SS_3)}^*(R_2) \cr
    &\times \exp\left[\left(-M_{\mathrm{cl}}
    -\frac{1}{2I} ~S(S+1) \right)
    \mathcal{T}  \right] =\exp\left[\left(-M_{\mathrm{cl}}
    -\frac{1}{2I} ~S(S+1) \right)
    \mathcal{T}  \right] \delta_{S'_3S_3},
\label{eq:68} 
\end{align}
where $R_1=R(\mathcal{T}/2)$ and $R_2=R(-\mathcal{T}/2)$.
Thus, the mass of the baryon with definite quantum numbers $S=T$ is
obtained to be
\begin{align}
M_{S=T} =   M_{\mathrm{cl}}+ \frac{1}{2I} ~S(S+1).
\label{eq:69}
\end{align}
Thus, the mass splitting between the nucleon $(S=T=1/2)$ and
$\Delta$ isobar $(S=T=3/2)$ is given as 
\begin{align}
    M_{\Delta-N}=M_{\Delta}-M_{N} = \frac{3}{2I},
\label{eq:70}
\end{align}
which is a well-known expression~\cite{Adkins:1983ya,
  Diakonov:1987ty}.

\begin{table}[h!]
\caption{\label{tab:2}%
Results for the moment of inertia. The first and
second columns list those of the valence- and sea-quark
contributions, whereas the third column represents the total
result. In the final column, The mass splitting of the $\Delta$ and
nucleon is given. } 
\centering
\renewcommand{\arraystretch}{2}
\setlength{\tabcolsep}{6pt}
\begin{tabular}{ccccc} \hline\hline
 & $I_{\rm val}~{\rm [fm]}$ & $I_{\rm sea}~{\rm [fm]}$ &
 $I~{\rm [fm]}$ & $M_{\Delta - N}~{\rm [MeV]}$  \\
\hline
  Present work &$1.0519$ & $0.3334$ & $1.3853$ & $213.67$  \\
  $\chi$QSM ($M=359$ MeV) &  $1.1733$ & $0.2367$ & $1.4100$ 
                                               & $209.93$\\ 
  $\chi$QSM ($M=420$ MeV) & $0.7752$ & $0.2553$ & $1.0306$ 
                                               & $267.62$\\  
\hline\hline
\end{tabular}
\end{table}  
In Table~\ref{tab:2}, we list the results for the moment
of inertia in unit of fm, and the $\Delta-N$ mass splitting in
comparison with those obtained from the $\chi$QSM, where the
momentum dependence of the dynamical quark mass is switched off. 
Before we proceed to the comparison, we will briefly explain the
results from the $\chi$QSM. When the momentum dependence of $M(k)$
turns off, a regularization must be introduced to tame the divergences
arising from quark loops. The proper-time or Pauli-Villars
regularization is often employed. The cutoff mass is fixed by
reproducing the experimental value of the pion decay constant
$f_\pi$. Since the single cutoff yields the current quark mass
$m_{\mathrm{u}}=m_{\mathrm{d}}\approx 15$ MeV to reproduce the pion
mass $m_\pi=140$ MeV, the double cutoff masses are often
introduced~\cite{Blotz:1992pw}. We will compare the
current results with those from the $\chi$QSM with the single cutoff
mass, since this is the simplest case. Though we take the chiral
limit in the present work, the comparison is still reliable, since the
current quark mass does not much contribute to the moments of
inertia. 

In the current theoretical framework, the dynamical quark mass at the
zero virtuality of the quark ($k^2=0$) is determined to be $M_0=359$
MeV, we choose the same value of $M_0$ in the $\chi$QSM, of which the
results are given in the third row. While the values of $I$ are almost
in agreement with each other, the current work produces a smaller
value of the valence-quark contribution, compared with that from the
$\chi$QSM. On the other hand, the present sea-quark contribution is
approximately 40~\% larger than that from the $\chi$QSM. This has
significant physical implications. The momentum dependence of the
dynamical quark mass originates from the fermionic zero-mode solutions
of the QCD equation of motion with the instanton background field. The
value of the dynamical quark mass emerges from the propagation of the
valence quark through the instanton medium, as explained in the
previous section. As the chiral symmetry is spontaneously broken by
this mechanism, pions arise as NG bosons. The presence of the $N_c$
valence quarks creates self-consistently the pion mean fields by
polarizing the vacuum. This implies that the pion mean fields are
originated from gluons. One can more clearly understand this unique
feature of the current theoretical framework from
Refs.~\cite{Diakonov:1995qy, Balla:1997hf}: gluon operators are
expressed in terms of the effective quark operators. These effective
quark operators can be further written in terms of the quark and pion
mean fields by bosonization. Very recently, Kim et
al.~\cite{Kim:2024cbq} showed that the twist-3 gluonic operator for
spin-orbit correlation can effectively be written in terms of the pion
mean field\footnote{Note that the effective gluonic operator derived
  in Ref.~\cite{Kim:2024cbq} is still a 
  local one with the momentum dependence of the dynamical quark mass
  turned off.}. This implies that the pion mean fields take their
origin from the gluon degrees of freedom. Since the momentum-dependent 
quark mass derived from the instanton vacuum is retained in
developing the current effective chiral theory of the nucleon,
properties of the instanton vacuum was also kept intact in the course
of the development. Thus, the large value of the sea-quark
contribution indicates the significant role of the instanton vacuum to
understand the internal structure of the nucleon. 

As shown in Eqs.~\eqref{eq:50} and~\eqref{eq:52}, the moments of
inertia include the nonlocal contributions that contain terms with the
derivatives of the form factor $F$ with respect to $\omega$ as defined
in Eq.~\eqref{eq:48}.  Had we ignored them, we would have the
following results: $I_{\mathrm{val}}=1.7496$ fm, 
$I_{\mathrm{sea}}=0.3154$ fm, and $I=2.0650$ fm. 
This would have given $M_{\Delta-N} = 143.34$ MeV, which is much
smaller than the current result and the experimental data. The
nonlocal terms make $I_{\mathrm{val}}$  reduced approximately by
40\%. On the other hand, $I_{\mathrm{sea}}$ is increased only by about
5\%. The final result for the moment of 
inertia is $I=1.385$ MeV, which yields $M_{\Delta}-M_N=213.67$
MeV. If we take the experimental center value of the $\Delta$ isobar,
we get $(M_{\Delta}-M_N)_{\mathrm{Exp}}=267.62$
MeV~\cite{ParticleDataGroup:2024cfk}. However, the $\Delta$ isobar is
known to have a large width $\Gamma_{\Delta}\approx 117$
MeV~\cite{ParticleDataGroup:2024cfk}, so the mass splitting of the
$\Delta$ and nucleon has large uncertainties. This leads to the fact
that its value lies in the wide range, i.e., $(200-400)$ MeV. Thus,
the current result $M_{\Delta}- M_N=214$ MeV is in qualitatively good
agreement with the experimental data. We also want to mention that the
$1/N_c$ meson-loop corrections are ignored in the current work, which
may have certain next-to-leading order corrections to the $\Delta-N$
mass difference.     

\section{Singly heavy baryons\label{sec:6}} 
A great advantage of the present theoretical framework is that both
light baryons and singly heavy baryons can be treated on an equal
footing. G.-S. Yang et al.~\cite{Yang:2016qdz} showed that singly
heavy baryons can be viewed as a bound state of the $N_c-1$ valence
quarks, which are bound by the pion mean fields created by the
presence of the $N_c-1$ valence quarks in a self-consistent
manner. The idea is rooted in heavy-quark spin-flavor
symmetry~\cite{Isgur:1989vq, Isgur:1991wq, Georgi:1990um}. In the
limit of the infinitely heavy quark mass ($m_Q\to \infty$), the spin
of the heavy quark inside a singly heavy baryon, $\bm{S}_H$, is
conserved. Consequently, the light-quark spin,
$\bm{S}_L=\bm{S}-\bm{S}_H$, is also conserved. This is called the
heavy-quark spin symmetry. Thus, the total spin of
the light quarks becomes a good quantum number, so that a singly heavy
baryon can be regarded as a bound state of a single heavy quark and a 
diquark. This allows one to classify singly heavy baryons in terms of
representations of the flavor SU(3) symmetry: $\bm{3}\otimes \bm{3} =
\overline{\bm{3}} \oplus \bm{6}$. While the baryon antitriplet
($\overline{\bm{3}}$) remains a spin singlet, the baryon sextet
($\bm{6}$) has both spin singlet and triplet. Being coupled to the
spin of the heavy quark, the singly heavy baryon in the 
$\overline{\bm{3}}$ has spin 1/2 whereas those in the $\bm{6}$ have
either spin 1/2 and spin 3/2. In the limit of $m_Q\to \infty$, the
baryon sextet with spin 1/2 and 3/2 are degenerate. This degeneracy
can only be removed by considering the $1/m_Q$ corrections. In the
isospin SU(2) symmetric case, we have only up and down quarks. Thus,
the singly heavy baryons can be decomposed as $\bm{\frac12}\otimes
\bm{\frac12} = \bm{0}\oplus \bm{1}$. Thus, the $\Lambda_c$ is 
 a isospin singlet, and $\Sigma_c$ and $\Sigma_c^*$ are isotriplet
 with spin 1/2 and 3/2, respectively.  

The $\chi$QSM was successfully applied to the
description of the singly heavy baryons: the mass splitting of the
low-lying singly heavy baryons~\cite{Yang:2016qdz}, their magnetic
moments~\cite{Yang:2018uoj, Kim:2018xlc, Kim:2019rcx} and radiative
decays~\cite{Yang:2019tst}, electromagnetic form
factors~\cite{Kim:2018nqf, Kim:2020uqo, Kim:2019wbg}, axial-vector
transitions~\cite{Suh:2022ean}, the quark spin
content~\cite{Suh:2022atr}, and so on. We briefly recapitulate how to
describe singly heavy baryons in the current framework. For detailed
formalism, we refer to Refs.~\cite{Kim:2019rcx,Kim:2018cxv,Kim:2021xpp}.
A singly heavy baryon state is defined as 
\begin{align}
|B,p\rangle &= \lim_{x_4\to-\infty} \exp(ip_{4}x_{4})
              \mathcal{N}(\bm{p}) \int d^3 x~
              \exp(i\bm{p}\cdot \bm{x}) (-i\Psi_h^\dagger(\bm{x}, x_4)
              \gamma_4) J_B^\dagger (\bm{x},x_4) 
              |0\rangle,\cr
\langle B,p| &= \lim_{y_4\to \infty} \exp(-ip'_4 y_4) 
               \mathcal{N}^*(\bm{p}') \int d^3 y~
              \exp(-i\bm{p}'\cdot \bm{y}) \langle 0| J_{B}
                 (\bm{y},y_4) \Psi_h (\bm{y},y_4),
\label{eq:71}
\end{align}
where $\Psi_h$ denotes the heavy quark field inside a singly heavy
baryon. The Ioffe-type current with the $N_c-1$ light valence quarks is 
expressed as 
\begin{align}
J_B(x) &= \frac1{(N_c-1)!} \epsilon_{\alpha_1\cdots \alpha_{N_c-1}} 
\Gamma_{(TT_3)(SS_3)}^{f_1\cdots  f_{N_c-1}}
\psi_{f_1 \alpha_1}(x)\cdots \psi_{f_{N_c-1} \alpha_{N_c-1}}(x),\cr  
J_{B}^\dagger(y) &= \frac1{(N_c-1)!} \epsilon_{\beta_1\cdots \beta_{N_c-1}}
\Gamma_{(TT_3)(SS_3')}^{g_1\cdots g_{N_c-1}}
(-i\psi^\dagger(y)\gamma_4)_{g_1\beta_1} \cdots
  (-i\psi^\dagger(y)\gamma_4)_{g_{N_c-1}\beta_{N_c-1}} .
\label{eq:72}
\end{align}
In the limit of $m_Q\to \infty$, we can express the heavy quark field
in a factorized form 
\begin{align}
\Psi_h(x) = \exp(-im_Q v\cdot x) \tilde{\Psi}_h(x),  
\label{eq:73}
\end{align}
where $\tilde{\Psi}_h(x)$ stands for a rescaled heavy-quark field
almost on mass-shell. It has no information on the heavy-quark mass in the
leading-order approximation in the heavy-quark expansion. $v$
represents the velocity of the heavy quark~\cite{Isgur:1989vq,
  Isgur:1991wq, Georgi:1990um,Shifman:1995dn}. The heavy-quark
propagator in the limit of $m_Q\to \infty$ is then obtained to be 
\begin{align}
G_h (y|x) = \left \langle y \left |   \frac1{\partial_t }\right |
  x\right \rangle= \Theta(y_4-x_4) \delta^{(3)}(\bm{y} - \bm{x}) ,    
\label{eq:74}
\end{align}
which indicates that the heavy quark inside a singly heavy baryon
remains as a static color source, and does not interact with the pion
background field\footnote{In fact, the heavy-quark propagator is
  expressed as an inverse of the covariant derivative, i.e. $G_h(y|x)=
 \langle y|(v\cdot D)^{-1} |x\rangle$. Since the gluon field in $D$
 can be rewritten in terms of the pion mean field, the heavy quark can
still interact with the pion mean field even in the limit of $m_Q\to
\infty$. This will be investigated in future works.}.  We obtain
the correlation function for the singly heavy baryon in the large
Euclidean time as 
\begin{align}
\langle J_B(y)\Psi_h(y)(-i\Psi_h^\dagger(x)\gamma_4)
  J_B^\dagger(x)\rangle_0 \underset{\mathcal{T} \rightarrow \infty}{\sim}~ 
\exp\left[-\left\{(N_{c}-1)E_{\mathrm{val}} +
  E_{\mathrm{sea}} +m_Q \right\} \mathcal{T}\right].
\label{eq:75}
\end{align}
Thus, the classical mass of the singly heavy baryon is expressed as 
\begin{align}
M_B = (N_c-1) E_{\mathrm{val}} + E_{\mathrm{sea}} + m_Q.  
\label{eq:76}
\end{align}

To find the profile function for the pion mean field in the presence
of the $N_c-1$ valence quarks, we need to modify the scalar and
pseusoscalar functions defined in Eq.~\eqref{eq:34} as 
\begin{align}
\tilde{S}(r) &:=  \cos\tilde{\Theta}(r) =-\frac{N_{c}-1}{2}z_{\rm
               val}\langle {\rm val} \vert 
\buildrel\leftarrow\over F(\gamma_{4})\buildrel
  \rightarrow\over F\vert {\rm val}\rangle
  +i\frac{N_{c}}{2}\int_{-\infty}^{\infty}
  \frac{d\omega}{2\pi}\sum_{n}\frac{\langle
  n\vert\buildrel \leftarrow\over F(\gamma_{4})
  \buildrel\rightarrow\over F\vert
  n\rangle}{\omega+i E_{n}}, \cr
\tilde{P}(r)&:= \sin\tilde{\Theta}(r) =-\frac{N_{c}-1}{2}z_{\rm val}\langle
              {\rm val}
  \vert\buildrel\leftarrow\over
  F(i\gamma_{4}\gamma_{5} \bm{\tau}\cdot
  \hat{\bm{n}})
  \buildrel\rightarrow\over F\vert {\rm
  val}\rangle
  +i\frac{N_{c}}{2}\int_{-\infty}^{\infty}
  \frac{d\omega}{2\pi}\sum_{n}\frac{\langle
  n\vert  \buildrel\leftarrow\over
  F (i\gamma_{4}\gamma_{5}\bm{\tau}\cdot
  \hat{\bm{n}})\buildrel\rightarrow
  \over F\vert n\rangle}{\omega+i E_{n}}.
\label{eq:77}
\end{align}
We expect that the pion mean field created by the presence of the
$N_c-1$ valence quarks will be weaker than that from the $N_c$ valence
quarks. 

\begin{figure}[htp]
\centering
\centering
\includegraphics[width=0.45\columnwidth]{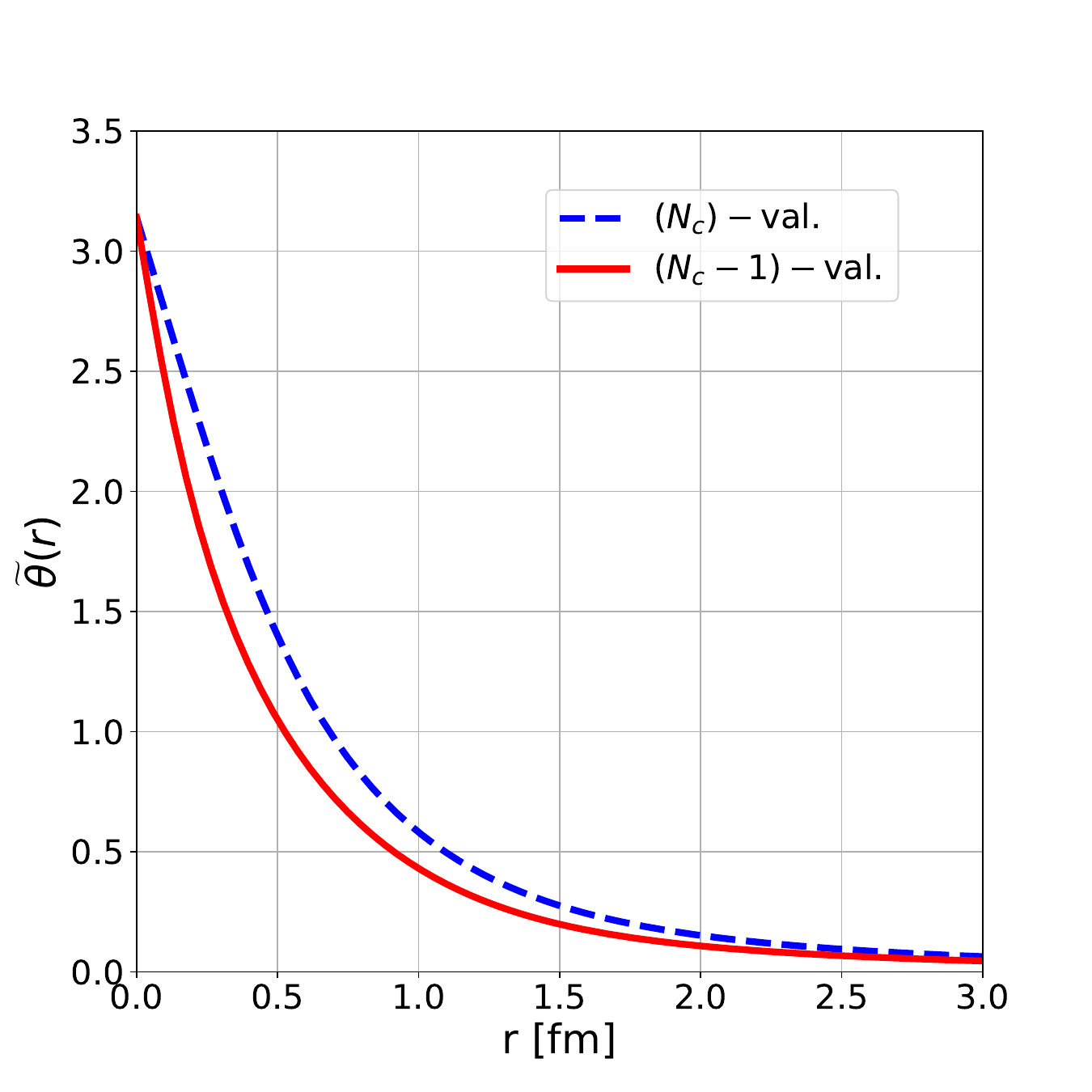}\;\;\;
\includegraphics[width=0.45\columnwidth]{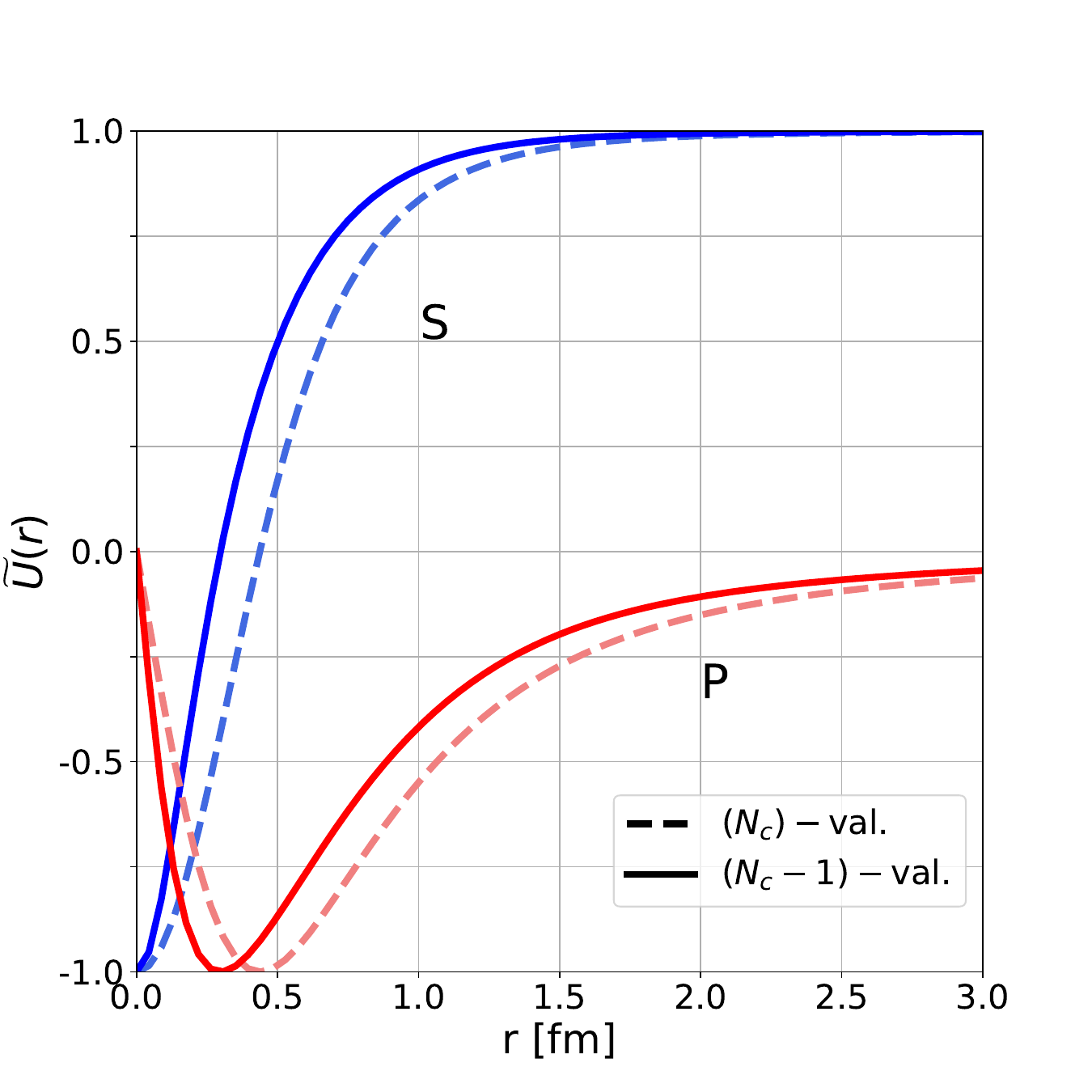}
\caption{\label{fig:4}%
The self-consistent profile function $\tilde{\Theta}(r)$ for
a singly heavy baryon is drawn in the left panel, whereas
$\tilde{S}(r)=\cos[\tilde{\Theta}(r)]$ and
$\tilde{P}(r)=-\sin[\tilde{\Theta}(r)]$ are depicted in the 
right panel. The results are compared with those for the nucleon. In
the figures, we use the same notation as in the case of the nucleon
for comparison.} 
\end{figure}

The left panel of Fig.~\ref{fig:4} depicts the result for the profile
function of the $N_c-1$ pion mean field. We compare it with that of
the $N_c$ one, which was already presented in Fig.~\ref{fig:3}. As
expected, the result is different from that of the $N_c$ one. As $r$ 
increases, the size of the profile function become narrower than that
of the $N_c$ one, which indicates that the classical singly heavy
baryon becomes more compact than the classical nucleon.
Consequently, the results for the $S(r)$ and $P(r)$ in
Eq.~\eqref{eq:77} show similar tendencies.
In Ref.~\cite{Kim:2018nqf}, it was demonstrated that the singly heavy
baryons are electromagnetically more compact than the
nucleon. Figure~\ref{fig:4} implies that the compactness of the singly
heavy baryon comes from a dynamical origin: as discussed above, the
$N_c-1$ mean field is weaker than the $N_c$ one, which brings about
the fact that the singly heavy baryons are generally more compact than
the nucleon. It also indicates that light quarks govern the dynamics
inside a singly heavy baryon.

\begin{table}[htp]
\caption{\label{tab:3}%
For the singly heavy baryons $(N_{c}-1)$-valence
quarks, the classical soliton mass $M_{\mathrm{cl}}$ split in
valence and sea contributions. } 
\centering
\renewcommand{\arraystretch}{2}
\setlength{\tabcolsep}{6pt}
\begin{tabular}{cccccc} \hline\hline
 & $E_{\mathrm{val}}~\mathrm{[GeV]}$ & $(N_c-1) E_{\mathrm{val}}~{\rm
[GeV]}$ & $E_{\mathrm{sea}}~{\rm [GeV]}$  & $M_{\mathrm{cl}}~{\rm
 [GeV]}$ \\   
\hline
Present work & $0.333$ & $0.666$ & $0.314$ & $0.980$  \\
$\chi$QSM ($M_0=420$ MeV)  & $0.329$ & $0.658$ & $0.382$ & $1.040$  \\ 
\hline\hline
\end{tabular}
\end{table}  
In Table~\ref{tab:3} we list the results for the valence- and
sea-quark energies. Let us first compare them with those for the
classical nucleon. Comparing them with those for the classical nucleon
given in Table~\ref{tab:1}, we find an interesting difference. In the
presence of the $N_c$ valence quarks, we obtain the valence-quark
energy $E_{\mathrm{val}}=257$ MeV, whereas we get
$E_{\mathrm{val}}=333$ MeV in the case of the $N_c-1$ valence
quarks. However, the total valence-quark energy for the $N_c-1$
soliton becomes larger than the classical nucleon. As discussed in
Ref.~\cite{Kim:2019rcx}, the gap equation does not exist when the
values of $M_0$ is approximately less than 420 MeV, so that the
$N_c-1$ pion mean field exists only for $M_0\gtrsim 420$ MeV. On
the other hand, we are able to produce the $N_c-1$ pion mean
field, even though we have $M_0=359$ MeV. This can be understood by
the same arguments discussed above. It is of great importance to keep
the momentum-dependent quark mass to retain the properties of the
instanton vacuum in producing the pion mean field. The classical mass
of the singly heavy baryon is given by Eq.~\eqref{eq:76}, so we get
approximately $M_{\Lambda_c} \approx 2.252$ GeV and $M_{\Lambda_b}
\approx 5.162$ GeV, which are comparable with both the masses of the
$M^{\rm exp.}_{\Lambda_c}=2.286$ MeV and $M^{\rm
  exp.}_{\Lambda_b}=5.619$, respectively.    

We can carry out the zero-mode quantization for the singly heavy
baryons. The procedure is exactly the same as the nucleon and delta
case. However, since the collective wave function for the quantized
$N_c-1$ soliton must be coupled to the heavy-quark spinor, we express
that for a singly heavy baryon as 
\begin{align}
\Psi_{(TT_3)(SS_3)} =\sum_{M_3=\pm 1/2} C_{S_Q M_3 S'S'_3}^{SS_3}
  \sqrt{2S'+1}(-1)^{T+T_3} D_{-T_3S'_3}^{S'=T} (R) \chi_{M_3},  
\label{eq:78}
\end{align}
where $S_Q$ and $M_3$ denote the spin and its third component for the
heavy quark, $S'$ and $S'_3$ represent those for the $N_c-1$ quark
system. $S$ and $S_3$ represent the spin and its third component of
a singly heavy baryon. Thus, the $\Lambda_c$ baryon belongs to the
representation $S'=T=0$ (isospinglet) and has spin $S=1/2$, whereas
the $\Sigma_c$ and $\Sigma_c^*$ baryons belong to the representation
$S'=T=1$ (isotriplet) and have $S=1/2$ and $S=3/2$, respectively. In
the absence of the $1/m_Q$ corrections, the masses of $\Sigma_c$ and
$\Sigma_c^*$ are degenerate. We can classify the bottom baryons in the
same manner. 

The corresponding mass of a singly heavy baryon is given by 
\begin{align}
M_B = M_{\mathrm{cl}} + m_Q + \frac1{2\tilde{I}} S(S+1),
\label{eq:79}
\end{align} 
where $\tilde{I}$ is the moment of inertia for the $N_c-1$ pion mean
field, which is expressed as 
\begin{align}
\tilde{T} = \tilde{I}_{\mathrm{val}} + \tilde{I}_{\mathrm{sea}}.  
\label{eq:80}
\end{align}
The expression for $\tilde{I}_{\mathrm{val}}$ is written as 
\begin{align}
\tilde{I}_{\mathrm{val}} &= 
\frac{N_{c}-1}{2}\sum_{n}\frac{z_{\mathrm{val}}}{E_{n}-E_{\mathrm{val}}}
\langle \mathrm{val}\vert t^{a} \vert n\rangle \langle n
\vert t^{a}\vert \mathrm{val}\rangle\big{\vert}_{\omega=-i
E_{\mathrm{val}}} -\frac{N_{c}-1}{8} z_{\mathrm{val}}\langle
\mathrm{val}  \vert T^{aa} \vert \mathrm{val}
\rangle \big{\vert}_{\omega=-i E_{\mathrm{val}}}
\label{eq:81}
\end{align}
The sea-quark contribution to $\tilde{I}$ is the same as that for
the $N_c$ pion mean field. However, since its value is changed by the
$N_c-1$ pion mean-field soluion.

\begin{table}[htp]
\caption{\label{tab:4}%
Results for the moment of inertia and its
valence- and sea-quark contributions, and the mass splitting 
$M_{\Sigma_c}-M_{\Lambda_c}$.} 
\centering
\renewcommand{\arraystretch}{2}
\setlength{\tabcolsep}{6pt}
\begin{tabular}{ccccc} \hline\hline
& $I_{\mathrm{val}}~{\rm [fm]}$ &  $I_{\mathrm{sea}}~{\rm
  [fm]}$ & $I_{T}~{\rm [fm]}$ & $M_{\Sigma_Q-\Lambda_Q}~{\rm [MeV]}$ \\   
\hline
Present work & 0.7650 & 0.1920 & 0.9570 & 206.20 \\
$\chi$QSM($M_0=420$ MeV) & 0.9591 & 0.1438 & 1.1029 & 178.92 \\ 
\hline\hline
\end{tabular}
\end{table}

Table~\ref{tab:4} presents the results for the moment of inertia for
the $N_c-1$ soliton and the mass difference between
$M_{\Sigma_Q-\Lambda_Q}$. The moments of inertia
demonstrate similar features as in the case of the $N_c$ soliton (see
Table~\ref{tab:3}). The result for the mass splitting between the
singly heavy baryons is given by $206.20$ MeV. Since we do not
consider the $1/m_Q$ corrections, the spin $1/2$ and $3/2$ heavy
baryons in the isospin triplet representation are
degenerate. Moreover, there is no difference between the charm and
bottom sector. The result $M_{\Sigma_Q-\Lambda_Q}$ is comparable with
the experimental data for the mass splittings of $\Sigma_Q$ and
$\Lambda_Q$ given by $M_{\Sigma_c-\Lambda_c}=167~{\rm MeV}$ and
$M_{\Sigma_b-\Lambda_b}=194~{\rm MeV}$. The present
result is in very good agreement with the data for the bottom
sector. Since we take the infinitely heavy-quark mass limit, it is
natural to describe the bottom baryon sector better than the charm
sector. 

We have checked if we can produce the $N_c-2$ pion mean field within
the present theoretical framework. In Ref.~\cite{Kim:2019rcx}, it was
shown that the $N_c-2$ pion mean field can only be created by using
values of $M_0$ larger than approximately $M_0\gtrsim 600$
MeV. Considering the fact that $M_0$ can be regarded as the coupling
strength between the quark fields and the pion mean field, we can
understand that it is unlikely to produce the $N_c-2$ pion mean field
within the current scheme. 

\section{Summary and conclusions}\label{sec:7}

In this work, we have developed an effective chiral theory for the
nucleon based on the low-energy effective QCD partition function
derived from the QCD instanton vacuum, retaining the momentum
dependence of the dynamical quark mass. We focused on the significant
physical implications of this momentum dependence, which originates
from the fermionic zero-mode solutions of the QCD equation of motion
in the instanton background field. The presence of the $N_c$ valence
quarks self-consistently creates the pion mean fields by polarizing
the vacuum. We obtained the classical nucleon mass
$M_{\text{cl}}=1.2680$ GeV, which is slightly larger than that
predicted by the local theory. Using zero-mode quantization, we
determined the moment of inertia $I=1.3853$ fm, from which we
predicted mass splittings: $M_{\Delta-N}=213.67$ MeV for the
$N$-$\Delta$ system and $M_{\Sigma_Q-\Lambda_Q}=206.20$ MeV for singly
heavy baryons. These results are in good agreement with experimental
data, particularly in the bottom sector where the heavy quark limit is
more applicable. This theoretical framework provides a unified
description of both light and singly heavy baryons, preserving the
properties of the instanton vacuum throughout its development. 
The results successfully reproduce experimental observations and
demonstrate the crucial role of the momentum-dependent dynamical quark
mass, implying that the pion mean fields originate from gluons. 

This work offers a consistent theoretical approach for exploring
gluonic observables in both light and heavy baryons, paving the way
for an improved understanding of baryon structure in the upcoming
Electron-Ion Collider (EIC) era. For singly heavy baryons, we are now
able to incorporate the covariant derivative, which can be interpreted
as the inverse of the static heavy-quark propagator. Since the gluon
field in the static propagator can be replaced by the pion mean field
via the QCD instanton vacuum, the heavy quark inside a singly heavy
baryon is no longer isolated from the surrounding light
quarks. Furthermore, gluons play a crucial role in addressing $1/m_Q$
corrections. This indicates that the current theoretical framework can
be used to derive $1/m_Q$ corrections and associated effective gluonic
operators. 
\medskip

\acknowledgments
HChK expresses his gratitude to C{\'e}dric Lorc{\'e} for invaluable
discussion and hospitality during his visit to Le Centre de Physique
Th{\'e}orique (CPHT) at {\'E}cole polytechnique, where part of the
present work was done. He is also grateful to the members of the CPHT
for their warm welcome. Y. W. Choi sincerely thanks H.-D. Son for the
fruitful discussions and encouragement, and H.-Y. Won for the useful
discussions on the chiral quark soliton model. The work was supported
by the Basic Science Research Program through the National Research
Foundation of Korea funded by the Korean government (Ministry of
Education, Science and Technology, MEST), Grant-No. 2021R1A2C2093368 
and 2018R1A5A1025563.     

\bibliography{NucleonIV}
\bibliographystyle{apsrev4-2}

\end{document}